\documentclass[aps,reprint,superscriptaddress]{revtex4-1}

\usepackage{hyperref}
\hypersetup{
     colorlinks   = true,
     citecolor    = red}
\usepackage{xr}
\usepackage{amsmath}
\usepackage{bm}
\usepackage{gensymb}
\usepackage[normalem]{ulem}
\usepackage{euscript}
\usepackage[pdftex]{graphicx}
\usepackage{xspace}
\usepackage{xfrac}
\usepackage{enumerate}
\usepackage{braket}
\usepackage{float,fancyvrb}
\usepackage[T1]{fontenc}
\usepackage{lmodern}
\graphicspath{{./Figures/}}
\usepackage{wrapfig}
\usepackage{scrextend}
\usepackage{comment}
\usepackage{url}
\usepackage{amssymb}
\usepackage{xcolor} 
\usepackage{lineno} 

\usepackage{mathrsfs}
\usepackage[scr=zapfc, scrscaled=1.15]{mathalfa}

\newcommand{\uu}[1]{\ensuremath{\, \mathrm{#1}}} 

\newcommand{\Bw}{{\vec{\Omega}}}

\newcommand{\Bt}{{\vec{\theta}}}
\newcommand{\BA}{{\vec{\mathscr{A}}}}
\newcommand{\dBt}{{\vec{\delta\theta}}}
\renewcommand{\d}{\mathrm{d}}
\newcommand{\e}{\mathrm{e}}
\def\abs#1{\left|#1\right|}
\def\bra#1{\langle#1|}
\def\ket#1{|#1\rangle}
\def\braket#1#2{\langle#1|#2\rangle}
\def\qexp#1#2{\bra{#2}#1\ket{#2}}

\def\tr#1{\mathrm{tr}\left[#1\right]}

\def\HCI{\hat{\mathcal{H}}_\mathrm{CI}}

\usepackage[colorinlistoftodos]{todonotes}

\begin{document}

\title{Exploring 2D synthetic quantum Hall physics with a quasi-periodically driven qubit}

\author{Eric Boyers}
\affiliation{Department of Physics, Boston University, 590 Commonwealth Ave., Boston, MA 02215, USA}
\author{Philip J. D. Crowley}
\affiliation{Department of Physics, Boston University, 590 Commonwealth Ave., Boston, MA 02215, USA}
\author{Anushya Chandran}
\affiliation{Department of Physics, Boston University, 590 Commonwealth Ave., Boston, MA 02215, USA}
\author{ Alexander O. Sushkov}
\email{email: asu@bu.edu}
\affiliation{Department of Physics, Boston University, 590 Commonwealth Ave., Boston, MA 02215, USA}
\affiliation{Department of Electrical and Computer Engineering, Boston University, Boston, MA 02215, USA}
\affiliation{Photonics Center, Boston University, Boston, MA 02215, USA}

\date{\today}

\begin{abstract}
Quasi-periodically driven quantum systems are predicted to exhibit quantized topological properties, in analogy with the quantized transport properties of topological insulators. We use a single nitrogen-vacancy center in diamond to experimentally study a synthetic quantum Hall effect with a two-tone drive. 
We measure the evolution of trajectories of two quantum states, initially prepared at nearby points in synthetic phase space. We detect the synthetic Hall effect through the predicted overlap oscillations at a quantized fundamental frequency proportional to the Chern number, which characterizes the topological phases of the system.
We further observe half-quantization of the Chern number at the transition between the synthetic Hall regime and the trivial regime, and the associated concentration of local Berry curvature in synthetic phase space.
Our work opens up the possibility of using driven qubits to design and study higher-dimensional topological insulators and semi-metals in synthetic dimensions.

\end{abstract}

\maketitle

\hypersetup{linkcolor=green}

Synthetic dimensions enrich the physical properties of low-dimensional systems and bring four and higher dimensional real-space models into the laboratory~\cite{boada2012quantum,celi2014synthetic,ozawa2016synthetic,yuan2018synthetic,ozawa2019topological}.
In particular, synthetic topological effects are intimately related to topological pumps in lower dimensions and time-independent topological systems in higher dimensions~\cite{thouless1983quantization,kraus2012topological,ozawa2016synthetic,thouless1983quantization,petrides2018six,lohse2018exploring}, and have a number of potential applications in quantum engineering~\cite{nayak2008non,goldman2016topological,barik2018topological,zhang2018topological,cooper2019topological}.
Such effects have so far been explored in several extended systems, ranging from cold atomic lattices to ring resonators and photonic waveguides~\cite{price2015four,price2017synthetic,Lohse2018,salerno2019quantized,tsomokos2010using,mei2016witnessing,ozawa2016synthetic,yuan2016photonic,yuan2018synthetic,lin2018three,price2019synthetic,Dutt2020,Zilberberg2018,ozawa2019topological,lustig2019photonic}.
However, designing and implementing a new generation of quantum devices that can take advantage of topological effects requires realizing synthetic topological states in well-controlled qubits with long coherence times and quantum architectures with scalability potential~\cite{Ma2018,shu2018observation,jelezko2004observation,buluta2011natural,clarke2008superconducting,dobrovitski2013quantum,kloeffel2013prospects,haffner2008quantum,devoret2004superconducting,langer2005long,taylor2005fault,trauzettel2007spin,gali2009theory,blatt2012quantum,harty2014high,wendin2017quantum,wang2017single,Martin2017,Crowley2019,crowley2019half}.

\begin{figure}[ht]
	\centering
	\includegraphics[width=.5\textwidth]{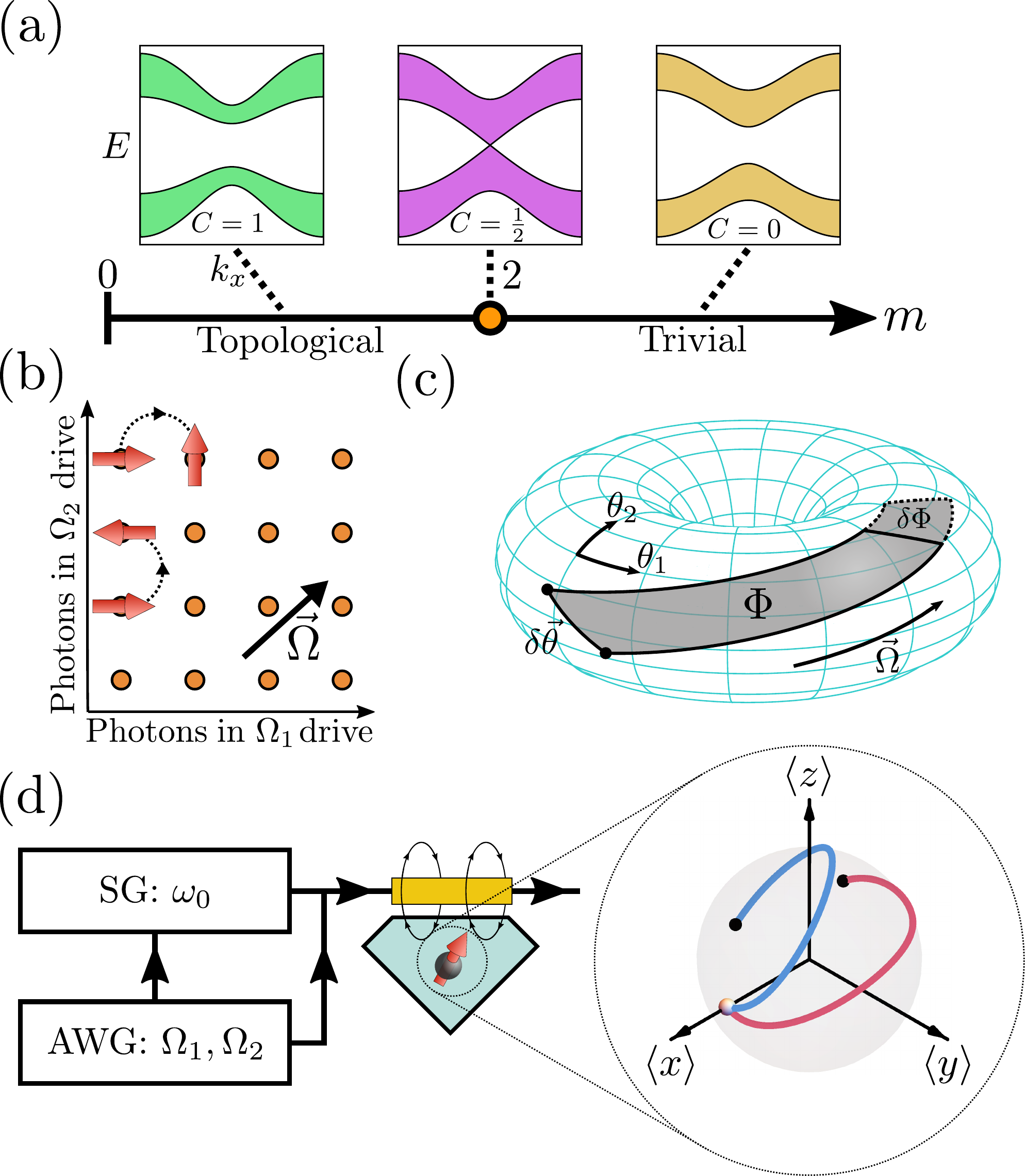}
	\caption{\textbf{(a)} Phase diagram and representative bandstructures of the half-BHZ model showing the topological $\left( 0 < m < 2 \right)$ and trivial $\left( m > 2 \right)$ regimes with ground state Chern numbers of $C=1$ and $C=0$ respectively, and the critical point $\left( m = 2 \right)$ with an effective Chern number of $C=\frac{1}{2}$ due to the Dirac point.  \textbf{(b)} Synthetic lattice formed by photon occupation numbers of each drive tone. Nearest-neighbor hops correspond to absorption or emission of a photon in one of the drive tones, while the effective electric field $\vec{\Omega}=(\Omega_1,\Omega_2)$ accounts for the photon energy in each drive. 
	\textbf{(c)} Synthetic Brillouin zone of drive phases $(\theta_1,\theta_2)$ conjugate to the synthetic photon lattice. The phases evolve linearly in time corresponding to dynamics under an effective electric field $\vec{\Omega}$. Two states, marked with black circles, are initialized with initial drive phases $\Bt_0$ and $\Bt_0 + \dBt$. After an evolution time, the states acquire a relative phase $\Phi$ due to the Berry curvature in the region bounded by their trajectories. Incrementing evolution time by $\delta t$ changes this phase by $\delta\Phi$. 
	\textbf{(d)} Experimental setup schematic. The amplitude of a carrier tone at frequency $\omega_0$ created by a signal generator (SG) is modulated by an arbitrary waveform generator (AWG), so that the RF fields oscillate at frequencies $\Omega_1$, $\Omega_2$ in the rotating frame. Signals are injected into a waveguide generating a RF magnetic field that drives the electronic spin of a single NV center in diamond. The spin Bloch sphere trajectories of the two states with initial drive phases $\Bt_0$ and $\Bt_0 + \dBt$ are shown as red and blue lines, diverging as they evolve and acquire a relative Berry phase.}
	\label{fig.seq}
\end{figure}

The Chern insulator is a band insulator that exhibits non-trivial topological features, such as a quantized Hall response when subject to weak electric fields.~\cite{thouless1982quantized,Hasan2010,bernevig2013topological,bansil2016colloquium}. 
The quantization is related to a topological invariant of the band structure known as the Chern number, which measures the integrated Berry curvature of a band.
A two-band model for a Chern insulator at half filling on a 2D square lattice is the half-Bernevig-Hughes-Zhang model (or the half-BHZ model)~\cite{bernevig2013topological,qi2006topological} with Bloch Hamiltonian $H(\vec{k}) = \sigma_x\sin k_x + \sigma_y \sin k_y - \sigma_z(m - \cos k_x - \cos k_y)$, where $\vec{k}=(k_x,k_y)$ is the Bloch momentum measured in units of the inverse lattice constant and $\sigma_i$ are the (pseudo-spin) Pauli matrices.
A Bloch state initialized in one of the bands will gain a Berry phase when the Bloch momentum varies adiabatically, tracing out a closed loop in the Brillouin zone.
When the path encloses the entire Brillouin zone, the Berry phase is quantized as an integer multiple of $2\pi$; the corresponding integer is defined to be the Chern number of the band. 
As the Chern numbers of the two bands sum to zero, we label the Chern number of ground (excited) band to be $+C$ ($-C$).
The parameter $m$ controls the phase diagram of the half-BHZ model at half filling (Fig.~\ref{fig.seq}(a)): when $0<m<2$, the model is in the Chern insulating phase with $C=1$, while for $m>2$, the model is a trivial insulator with $C=0$. At the transition at $m=2$, the model is semi-metallic with a Dirac point in the band structure.

We experimentally study the band topology of a half-BHZ model with a single spin-$\tfrac12$ effective qubit, formed by the electronic spin of a nitrogen-vacancy (NV) center in diamond. Using a magnetic resonance approach, we drive the NV spin qubit with radiofrequency (RF) magnetic fields at two incommensurate frequencies $\Omega_1,\Omega_2$. The role of the components of the crystal momentum $\vec{k}=(k_x,k_y)$ is then played by the drive phases $\vec{\theta}_t = \left(\theta_{t1}, \, \theta_{t2}  \right) = (\Omega_1,\Omega_2)t+(\theta_{01},\theta_{02})$, where $\theta_{01},\theta_{02}$ are the initial drive phases. 
For drive Rabi frequency $\gamma B_0$, where $\gamma$ is the NV center gyromagnetic ratio and $B_0$ is the RF field amplitude, our system is described by the single-spin Hamiltonian
\begin{align}
    H(\vec{\theta}_t) = \gamma B_0& \left[\sigma_x \sin \theta_{t1} + \sigma_y \sin \theta_{t2} \right. \nonumber \\
    &\left. - \sigma_z (m - \cos \theta_{t1} - \cos \theta_{t2}) \right], \label{eq:CI_t}
\end{align}
which is obtained from the half-BHZ Bloch Hamiltonian by replacing $\vec{k}$ with $ \vec{\theta}_t$ ($\hbar=1$ throughout).
The incommensurate nature of the drives causes the system to evolve quasiperiodically in time~\footnote{For the drives to be incommensurate on an experimental time-scale $t_\mathrm{exp}$, we need $\Omega_2/\Omega_1$ be distinguishable from all rationals $p/q$ that imply an overall period $T := 2 \pi q / \Omega_1 < t_\mathrm{exp}$. See ~\cite{supplement} for details.}, as the phases $\theta_{t1},\theta_{t2}$ sample the synthetic Brillioun zone torus $[0,2\pi)^2$ uniformly. To explore the topology of the single-particle bands, we enforce adiabaticity and follow the time evolution of the instantaneous eigenstates of the Hamiltonian (\ref{eq:CI_t}). We emphasize that there is no notion of a chemical potential in our synthetic model.

Fourier transformation of the time-dependent Schr\"odinger equation with Hamiltonian (\ref{eq:CI_t}) yields the half-BHZ tight-binding model with an additional term $\vec{n}\cdot\vec{\Omega}$, where $\vec{n}=(n_1,n_2)$ denotes the photon numbers in the two drive tones~\cite{shirley1965solution,sambe1973steady,ho1983semiclassical,verdeny2016quasi,peng2018topological,peng2018majorana,Martin2017,Crowley2019,crowley2019half,nathan2019topological}. Equivalently, interpreting the term $\vec{n} \cdot \vec{\Omega}$ as the potential energy of a uniform electric field $\vec{\Omega}$, the synthetic model corresponding to eq.~\eqref{eq:CI_t} is the half-BHZ tight-binding model in the presence of a uniform electric field. The photon numbers provide the synthetic dimensions. Nearest-neighbour hops on the synthetic lattice correspond to absorption or emission of a single photon into one of the drives, while the term $\vec{n} \cdot \vec{\Omega}$ accounts for the energy of the photons in each of the drives, fig.~\ref{fig.seq}(b). In the topological phase $0<m<2$ and in the adiabatic limit $|\vec{\Omega}| \ll \gamma B_0$, this model exhibits a long-lived quantized Hall effect that manifests as an energy current $\Omega_1 \Omega_2 C/(2\pi)$ between the two drives~\cite{Martin2017,Crowley2019}.

The most direct way to probe the topological properties of the qubit dynamics is to measure the Berry curvature, which can be extracted from the geometric phase accumulated as the system is adiabatically driven around a closed path in the synthetic Brillioun zone.
This experiment would be difficult to perform in a Chern insulating material due to the required precise control of wavepackets and low scattering from impurities. Here we demonstrate this approach with a well-controlled driven qubit platform, using only projective qubit measurements. 
It is natural to use the effective electric field $\vec{\Omega}$ to generate a path in the Brillouin zone: evolving for time $t$ generates a phase translation by vector $\vec{\Omega} t$, and subsequently applying a small perturbation $\dBt$ to the drive phase without altering the spin state, evolving backwards in time, and applying the inverse phase perturbation yields the closed path in fig.~\ref{fig.seq}(c). While traversing this path, each eigenstate accumulates a dynamical phase and a geometric phase. As the dynamical phase acquired during the forward and backward evolutions cancel to leading order in $\dBt$, the geometric contribution dominates. The geometric phase $\Phi$ is the Berry curvature integrated over the area $A$ bounded by the path:
\begin{equation}
    \Phi_j = \int_{A} d^2 \vec{\theta} \, \mathscr{B}_j (\vec{\theta}),
\end{equation}
where $\mathscr{B}_j (\vec{\theta}) = 2 \operatorname{Im}  \braket{\partial_{\theta_2}\phi^j}{\partial_{\theta_1}\phi^j} $ is the Berry curvature~\cite{berry1984quantal}, and $\ket{\phi^j(\vec{\theta})}$ for $j=1,2$ are respectively the ground and excited (instantaneous) eigenstates of the Hamiltonian $H(\vec{\theta})$.
At short times, the growth of $\Phi_j$ is set by the local Berry curvature $\partial_t \Phi_j(t) \approx | \dBt \times \vec{\Omega} | \mathscr{B}_j(\vec{\theta_t})$, which varies across the Brillioun zone. On longer evolution time scales, the drive phase uniformly samples the Brillouin zone and $\Phi_j(t)$ increases at an approximately constant rate set by the average value of the Berry curvature in band $j$. As the average Berry curvature is proportional to the Chern number, we obtain:
\begin{align}
\Phi_j(t) \approx  (-1)^{j+1}| \dBt \times \vec{\Omega} | \frac{C t} {2\pi} .
\end{align}
The average Berry curvature of the ground and excited state bands are equal and opposite.
For a spin initialized in a superposition of the instantaneous ground and excited states at $t=0$, the geometric phase $\Phi$ thus appears as a relative phase shift between the spin states and is given by:
\begin{align}
    \Phi(t) \equiv \Phi_1(t) - \Phi_2(t) = | \dBt \times \vec{\Omega} | \frac{C t}{\pi} \label{Eq:Phit}
\end{align}
In our approach this spin rotation is measured by determining the overlap between the final and the initial spin states. For a more general derivation of eq.~\eqref{Eq:Phit} away from the adiabatic limit, see the supplementary materials~\cite{supplement}.

We implement the synthetic half-BHZ Hamiltonian \eqref{eq:CI_t} in the rotating frame of the NV center electronic spin. A signal generator addresses the $\ket{m_s=0} \leftrightarrow \ket{m_s=+1}$ spin-flip transition at its resonant carrier frequency $\omega_0$, creating an effective spin-1/2 system, fig.~\ref{fig.seq}(d). The $x$- and $y$-components of the driving field are produced by independent channels of a waveform generator and implemented as the $x$- and $y$-quadratures of the carrier signal. The $z$-component of the driving field is an output of a separate generator~\cite{supplement}. To avoid hyperfine effects due to the nitrogen nuclear spin, we perform all experiments with a static external magnetic field tuned to the NV excited state level anti-crossing, where optically pumping the NV center polarizes both the NV electronic and nuclear spins~\cite{Jacques2009}.

In order to study the topological properties of the lower band of the synthetic Chern insulator system, we must ensure that the spin qubit evolution follows the instantaneous eigenstates of the Hamiltonian \eqref{eq:CI_t} without excitation. At finite drive frequencies, the qubit will eventually excite due to Landau-Zener transitions between the instantaneous eigenstates, and the topological phase will last only for a finite `pre-thermal' evolution time~\cite{zener1932non,landau1937theory,landau1937theory2,de2010adiabatic,Crowley2019}. We suppress these diabatic excitations by adding a counter-diabatic potential $V_{\mathrm{CD}}$ to the Hamiltonian~\cite{demirplak2003adiabatic,berry2009transitionless,del2013shortcuts,sels2017minimizing}. Since our model is a two level system, we calculate and implement $V_{\mathrm{CD}}$ exactly, see supplementary materials for further details~\cite{supplement}.

Closed-path trajectory measurements could be implemented by a sequence of appropriate phase shifts of the drive tones. However, in practice, it is easier to measure the time evolution of two quantum states prepared in the same superposition spin state at $t=0$,
\begin{align}
    \ket{\psi(0)} = \ket{\psi'(0)} = \frac{(\ket{\phi^1(\Bt_0)}+\ket{\phi^2(\Bt_0)})}{\sqrt{2}},
\end{align}
but at different initial drive phases $\Bt_0$ and $\Bt_0+\dBt_0$, Fig.~\ref{fig.seq}(c). For example, with $m=1$, $\Bt_0=\left( 0, \pi/2 \right)$,  the instantaneous eigenstates $\ket{\phi^j(\vec{\theta})}$ are $\ket{\pm y}$ so that $\ket{\psi(0)} = \ket{\uparrow}$. Since the drive phase perturbation does not alter the spin state, the geometric phase arises entirely from the quantum state evolution under the effective electric field $\vec{\Omega}$. Therefore, after evolving for time $t$, the two states acquire a relative phase $\Phi(t)$ equal to that acquired by a single quantum state driven in a closed path (eq.~\eqref{Eq:Phit}). We determine this phase from the quantum state overlap~\cite{supplement} 
\begin{align}
    F(t) &= \left| \braket{\psi'(t)}{\psi(t)} \right|^2 
    = \cos^2 \left( \frac{\Phi(t)}{2}   \right)  \nonumber \\
    &= \tfrac12 \big( 1 + \langle \sigma_x \rangle \langle \sigma_x \rangle' + \langle \sigma_y \rangle \langle \sigma_y \rangle' + \langle \sigma_z \rangle \langle \sigma_z \rangle'  \big),
    \label{eq:Chern_def}
\end{align}
where $\langle \sigma_i \rangle = \braket{\psi(t)| \sigma_i }{\psi(t)}$. Following state preparation and evolution for variable time $t$, we use a suitable $\pi/2$ pulse to rotate the spin state and measure one of these spin projections by NV center fluorescence detection. We repeat each of these measurements $\approx 10^6$ times, averaging the results to construct the state overlap. Quantum state evolution is interleaved with the spin echo protocol to improve spin qubit coherence. Spin dynamics are well-described by the synthetic half-BHZ model up to timescales on the order of spin dephasing time, which for this NV center is measured to be $T_2=(125\pm 7)$~$\mu s$. We model decoherence with an exponential decay of spin expectation values, so that the overlap saturates at $0.5$ for long evolution times~\cite{supplement}.

\begin{figure}[t]
	\centering
	\includegraphics[width= 1\textwidth]{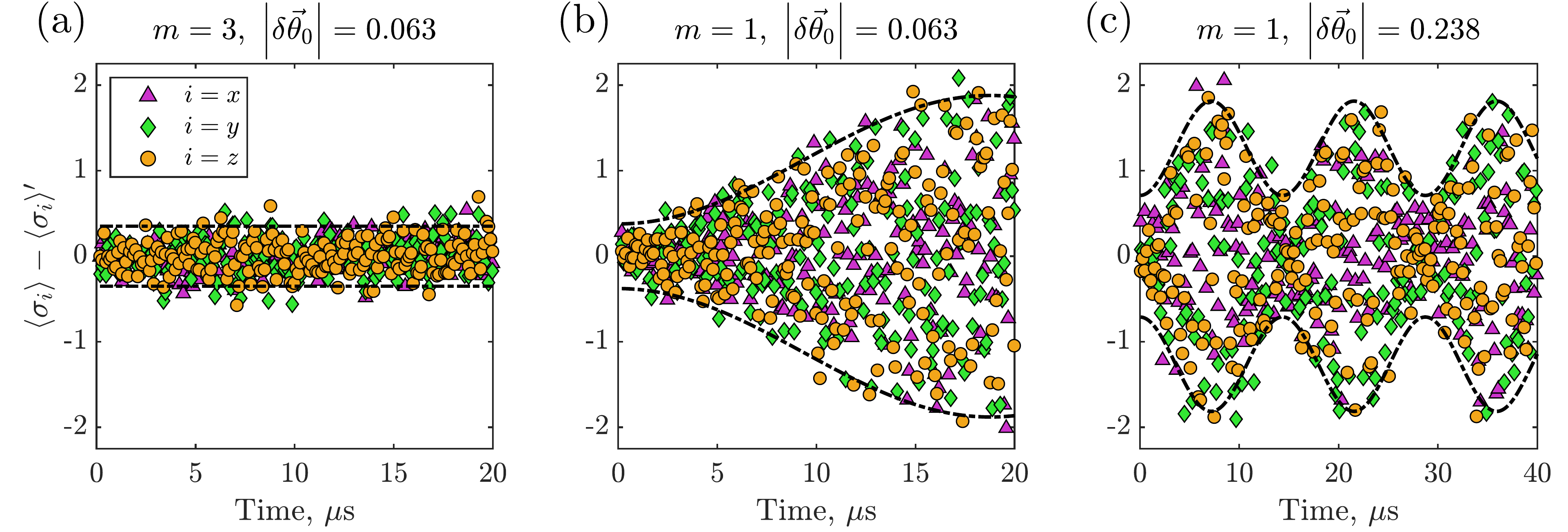}
	\caption{Measurements of spin projection evolution for trajectories starting with different initial drive phases and the same initial spin state $\ket{\psi(0)} =(\ket{\phi^1(\Bt_0)}+\ket{\phi^2(\Bt_0)})/\sqrt{2}$. All plots show on y-axis the difference in expectation values, $\langle \sigma_i \rangle - \langle \sigma_i \rangle'$ ($i=x,y,z$), and dashed lines are envelopes serving to guide the eye. Experimental parameters values: $\gamma B_0 = 2\pi \times 0.25 \uu{MHz}, \, \vec{\Omega}=2\pi \left(0.5, 0.5 \varphi \right) \uu{MHz}, \, \Bt_0 = \left(0, \pi/2 \right), \, \dBt_0 = \left(\varphi/30, -1/30 \right)$, with $\varphi$ equal to the golden ratio so that the drive frequencies are incommensurate.
	\textbf{(a)} Measurements for the system in the trivial phase ($m=3$): trajectories of the two states stay close to each other.
	\textbf{(b)} Measurements for the system in the topological phase ($m=1$): trajectories of the two states diverge as the states acquire increasing geometric phase.
    \textbf{(c)} Measurements for the system in the topological phase ($m=1$) with larger initial distance $\dBt_0 = \left(\varphi/8, -1/8 \right)$ and longer evolution time: the states periodically diverge and rephase.}
	\label{fig.divergence.diff}
	
\end{figure}

Topological features of spin dynamics are already apparent in the measurements of the spin projection expectation values after variable evolution time $t$.
In the trivial regime ($m > 2$), the geometric phase is zero since $C=0$ in both bands and no relative phase accumulates between $\ket{\psi(t)}$ and $\ket{\psi'(t)}$. Indeed the quantum states remain close to each other throughout their evolution and differences between spin expectation values are small, fig.~\ref{fig.divergence.diff}(a).
In the topological regime ($0 < m < 2$), the behaviour is markedly different. The states $\ket{\psi(t)}$, $\ket{\psi'(t)}$ diverge from each other because they acquire a Berry phase difference that increases with evolution time. 
This divergence is seen in fig.~\ref{fig.divergence.diff}(b) as the increasing divergence of spin projection measurements. 
Increasing the initial phase perturbation between the states $\ket{\psi(0)}$, $\ket{\psi'(0)}$ increases the rate of divergence and allows us to observe overlap oscillations as the total geometric phase becomes multiples of $2\pi$, fig.~\ref{fig.divergence.diff}(c).

\begin{figure}[ht]
	\centering
	\includegraphics[width= 1\textwidth]{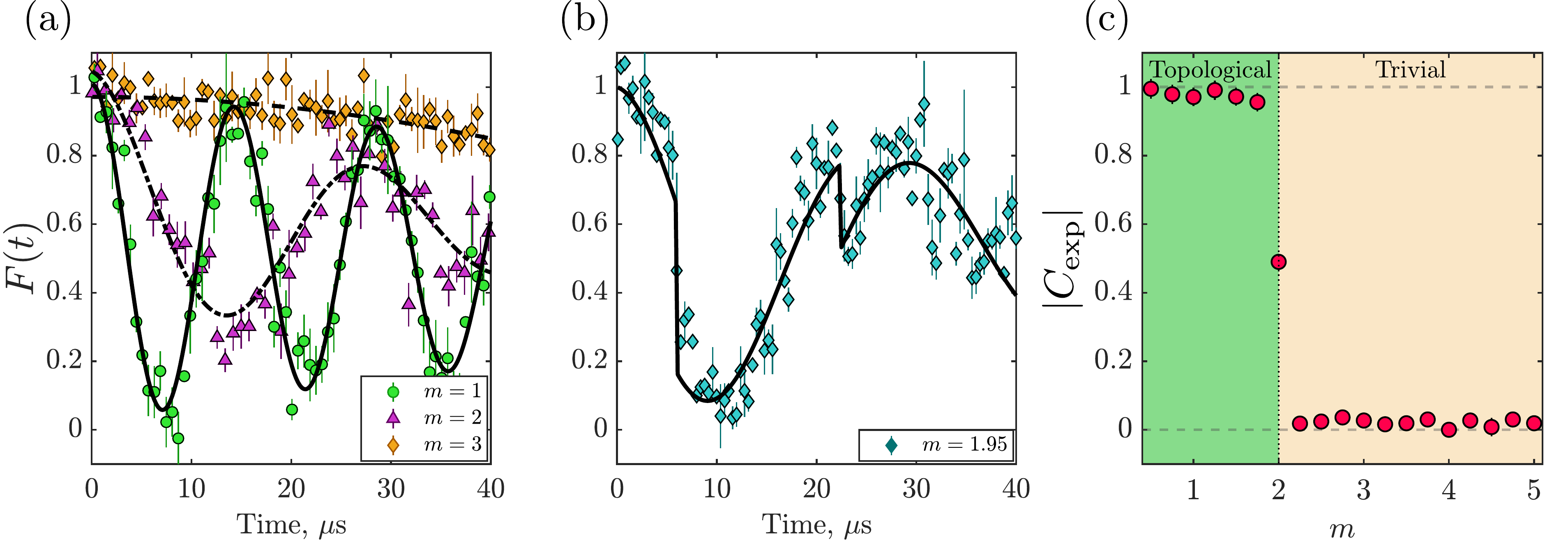}
	\caption{\textbf{(a)} Evolution of quantum state trajectories measured by their overlap $F(t)$. In the topological phase ($m=1$, green circles, solid line fit), the overlap oscillates with frequency proportional to the Chern number. In the trivial phase ($m=3$, orange diamonds, dashed line fit), the trajectories remain close at all times, and the overlap slowly decays due to qubit decoherence. At the critical point ($m=2$, purple triangles, dot-dashed line fit), the overlap oscillates at half the frequency of the topological regime. Fits are performed using the model $ F(t) = 1/2 + \left( 1/2 -  A \sin^2 \omega_{\text{fit}}t \right) e^{-t/\tau} + \delta$ where $\omega_{\text{fit}} = \frac{\abs{ \vec{\Omega} \times \dBt_0}}{2 \pi} \left| C_\textrm{exp} \right|$. Data are binned for clarity. Error bars represent 1 standard error and may be smaller than data markers. Experimental parameters are the same as in Figure \ref{fig.divergence.diff}(c): $\gamma B_0 = 2\pi \times 0.25 \uu{MHz}, \, \vec{\Omega}=2\pi \left(0.5, 0.5\varphi \right) \uu{MHz}, \, \Bt_0 = \left(0, \pi/2 \right), \, \dBt_0 = \left(\varphi/8, -1/8 \right)$. 
	\textbf{(b)} The overlap near the critical point, $m=1.95$, for the same parameters as in a. Sudden changes in overlap occur when either of the two states pass near the point $\Bt = \left( 0, \, 0 \right)$ around which the Berry curvature is concentrated. \textbf{(c)} The phase diagram of the half-BHZ model implemented with the NV center spin qubit. Red data points show Chern number measurements. Parameters are the same as in (a), error bars are of the order of data marker size and may be obscured.}
	\label{fig.divergence.fid}
\end{figure}

In order to quantify the Chern number of the synthetic BHZ model across its phase diagram, we measure $F(t)$ over long state evolution times, when the system path covers a large fraction of the Brillouin zone, fig.~\ref{fig.divergence.fid}(a).
We extract the absolute value of the Chern number of the instantaneous ground state band $|C_\textrm{exp}|$ from a fit to the overlap oscillation frequency predicted by eqs.~\eqref{eq:Chern_def},~\eqref{Eq:Phit}. The measured absolute values of the Chern number at several values of $m$ are:
\begin{align}
    |C_\textrm{exp}|  \left\{ \begin{array}{cc}
   =  0.97 \pm 0.03, & m=1 \textrm{ (Topological)} \\
   \leq 0.042, & m=3 \textrm{ (Trivial)} \\
  = 0.50 \pm 0.02, & m=2 \textrm{ (Critical)}
   \end{array} \right.
\end{align}
with all uncertainties and limits reported at the $1\sigma$ level. The best-fit value in the topological phase is in good agreement with the theoretical prediction of $C=1$. In the trivial phase ($m = 3$), no overlap oscillations are visible, consistent with the $C=0$ prediction. The observed overlap decline is consistent with qubit decoherence, although a conservative assumption that the decline is entirely caused by a small geometric phase results in an upper limit on the Chern number in the trivial phase: $\left| C_\textrm{exp} \right| \leq 0.042$.

The behavior near the critical point ($m = 2$) is more complex. As $m \to 2$ from below, a nascent Dirac point forms around $\Bt=(0,0)$, and half the Berry curvature concentrates in this region~\cite{bernevig2013topological,crowley2019half}. The remaining half of the Berry curvature remains distributed throughout the torus. Away from $\Bt=(0,0)$, trajectories sample only this distributed component, and thus the phase difference $\Phi(t)$ grows at half the rate of the topological regime. When the trajectories pass close to the nascent Dirac point, the states sample the concentrated component causing $\Phi$ to rapidly change. These sudden jumps are visible in the state overlap in Fig.~\ref{fig.divergence.fid}(b) at evolution times $t = 6\mu\text{s}, \, 24\mu\text{s}$. At the critical point $m = 2$, the concentrated component of the Berry curvatures sharpens to the Dirac point, and the phase jumps in $\Phi$ are of value $2 \pi$. These steps do not show in the overlap, resulting in smooth oscillations with half the frequency of the topological phase, Fig.~\ref{fig.divergence.fid}(a).
Indeed, the best-fit value of the Chern number at the critical point is $\left| C_\textrm{exp} \right| = 0.50 \pm 0.02$, consistent with the theoretical prediction~\cite{crowley2019half}.

In order to fully explore the topological phase diagram of our model, we perform the overlap measurements and extract Chern numbers for a range of values of $m$ with fixed initial phase $\vec{\theta}_0 = (0, \pi/2)$, Fig.~\ref{fig.divergence.fid}(c). In agreement with theory, we find two phases with ground state Chern numbers $C=0$ and $C=1$. Experiments at different initial drive phase $\Bt_0$ show that the extracted Chern number has weak dependence on $\Bt_0$ due to a transient effect~\cite{supplement}. Averaging over the initial phases yields $\left< \left| C_\textrm{exp} \right| \right>_{\Bt_0} = 0.997 \pm 0.007$ at $m = 1$ and $ \left< \left| C_\textrm{exp} \right| \right>_{\Bt_0} \leq 0.025$ at $m = 3$, consistent with the theoretical prediction.

\begin{figure}[ht]
	\centering
	\includegraphics[width= .5\textwidth]{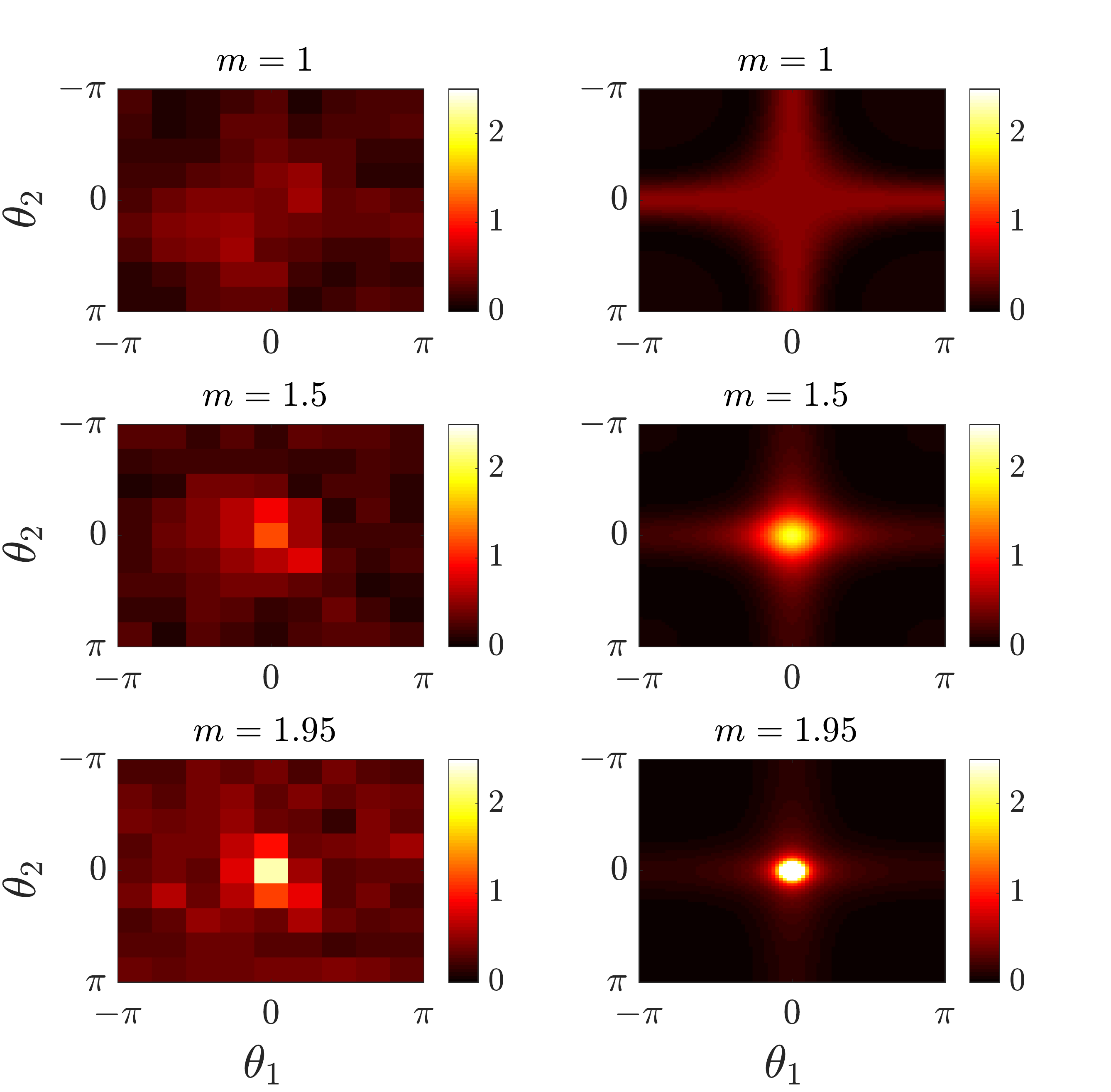}
	\caption{Local Berry curvature in the synthetic Brillioun zone for three values of $m$ approaching the critical point at $m=2$ from below. Color scale represents the magnitude of the local Berry curvature, which concentrates near the point $\Bt = \left( 0, \, 0 \right)$, exhibiting a Dirac point at $m=2$. Experimental measurements (left) are consistent with the exact Berry curvatures (right). Parameters are the same as in fig.~\ref{fig.divergence.fid} except for values of $m$ as noted. }
	\label{fig.berryMap}
\end{figure}

In addition to exploring the global topological features of the model, we also use the overlap measurement to investigate local features of the Berry curvature over the synthetic Brillouin zone. Since the Berry curvatures of the ground and excited bands have equal magnitude but opposite sign, a small increment of state evolution time $\delta t$ yields a change in the relative phase between the states of $\delta\Phi \approx  2 \left| \dBt \times \vec{\Omega} \delta t \right| \mathscr{B}_1(\Bt) $, Fig~\ref{fig.seq}c. Therefore the time derivative of geometric phase is proportional to the local Berry curvature $\mathscr{B}_1(\Bt)$. Directly using experimental data to evaluate the derivative $\d F / \d t=-(\frac{1}{2}\sin\Phi)\d\Phi/\d t$ introduces spurious modulation, caused by the $\sin\Phi$ pre-factor. 
A more robust approach is to average measurements of $(\d F/\d t)^2$ over a large set of evolution paths that start from different initial drive phases $\Bt_0$, but end at the same point $\Bt_\mathrm{f}$ in the Brillouin zone. The pre-factor then averages to $1/2$, resulting in:
\begin{align}
    \label{eq:fidelDerivRMS}
    \left< \left( \frac{dF(\Bt_\mathrm{f})}{dt} \right)^2 \right>_{\Bt_0} = \frac{1}{2} \left| \dBt \times \vec{\Omega}   \right|^2 \mathscr{B}_1^2(\Bt_\mathrm{f})
\end{align}
We use the above relationship and overlap data to obtain the local Berry curvature across the entire Brillouin zone for three different values of parameter $m$, fig.~\ref{fig.berryMap}. Our measurements show that the local Berry curvature increases and concentrates near the minimum energy gap at $\Bt = \left( 0, \, 0 \right)$ in the vicinity of the critical point at $m=2$. This concentration accounts for the overlap jumps observed in fig.~\ref{fig.divergence.fid}(b), and heralds the formation of the Dirac point at the topological phase transition. Our approach to measuring local Berry curvature is closely related to that used in cold atomic systems in optical lattices~\cite{flaschner2016experimental}, although the observation of a local Berry curvature in materials still poses a challenge~\cite{schuler2020local}.

In conclusion, we observe a synthetic 2D quantum Hall effect and a synthetic Dirac point by measuring their associated integer and half-integer quantized Chern numbers. In the topological regime, the two-tone driven qubit mediates a transient average energy current between the drives given by $P_{1\to 2} = \Omega_1 \Omega_2 C/(2\pi)$. This effect may find applications in next-generation quantum devices such as frequency converters, in state preparation and cooling of quantum cavities, and in quantum metrology~\cite{Martin2017,Crowley2019}. Although the energy current is long-lived under adiabatic conditions ($|\Omega| \ll \gamma B_0$), its value is small for a single-qubit system. Our work could be extended to make the energy current experimentally accessible by amplification using driven qubit ensembles and by implementing our counter-diabatic protocols to increase the lifetime of the current away from the adiabatic limit. Additionally, our work opens up a route to studying higher-dimensional topological systems in qubits by applying more mutually incommensurate tones, and demonstrates that overlap oscillations are a general tool for characterizing such systems.

The authors thank O.~P.~Sushkov for useful discussions.
The authors acknowledge support from the Alfred P. Sloan foundation through the grant FG-2016-6728 (E.B. and A.S.) and a Sloan research fellowship (A.C. and P.C.). This work was supported by NSF DMR-1752759 (A.C. and P.C.).

\bibliography{library_e112,additional_bib}

\onecolumngrid

\clearpage

\clearpage

\clearpage

\clearpage

\clearpage

\twocolumngrid

\appendix

\section{Theory of quasi-periodically driven quantum systems and the topological origin of overlap oscillations}

In the main text we gave an intuitive explanation of how quantum state overlap oscillations, Eq.~\eqref{eq:Chern_def}, arise due to a topological classification in the adiabatic limit. In this supplementary section we show that (i) these oscillations can arise even far from the adiabatic limit and (ii) are evident in the behaviour of generic observables, and thus do not require the reconstruction of the quantum state overlap (though the quantum state overlap oscillations indeed provides a particularly clear signature). To this end, we adopt the formalism of the topological classification of quasi-periodically driven few-level systems from Ref.~\cite{Crowley2019} and re-derive Eq.~\eqref{eq:Chern_def} for the oscillations in the quantum state overlap. 


This section proceeds as follows: in Sec.~\ref{app.Theory.back} we cover background on quasiperiodically varying functions; Sec.~\ref{app.QES} then obtains the steady states of quasi-periodically driven systems; Sec.~\ref{app.topoc} explains their topological classification; and, finally, Sec.~\ref{app.SOO} derives the quantum state overlap oscillation in this context, and relates this to previous work.

\subsection{Background on quasi-periodicity}
\label{app.Theory.back}

We begin by recounting some elementary properties of quasi-periodically varying functions. A more complete account of this topic is provided in standard introductory texts on Diophantine approximation, e.g. Refs.~\cite{casselsintroduction,hindry2013diophantine}.

Consider a two-level quantum system driven by a Hamiltonian with explicit quasi-periodic time dependence. Such a Hamiltonian can be written as
\begin{align}
    H_t &= H(\Bt_t)
    \\
    \Bt_t &= \Bt_0 + \Bw t
\end{align}
where $H(\Bt_t)$ has continuous periodic dependence on each phase angle in the vector $\Bt_t = (\theta_{t1},\theta_{t2})$, and the frequencies in $\Bw = (\Omega_1,\Omega_2)$ are chosen to be rationally independent,
\begin{equation}
    \frac{\Omega_2}{\Omega_1} = \beta \not\in \mathbb{Q}.
    \label{eq:not_rat}
\end{equation}
Quasi-periodic functions do not repeat in time ($\Bt_t = \Bt_0$ if and only if $t = 0$). However, they almost repeat after ``almost periods'' $T_n$. Almost periods correspond to times when $\Bt_t$ comes closer to its initial value $\Bt_0$ than it has done at any point along the trajectory previously. Thus, the sequence $\Bt_{T_n}$ converges to $\Bt_0$ as $n \to \infty$:
\begin{equation}
    \lim_{n \to \infty} \Bt_{T_n} = \Bt_0.
\end{equation}
The $n$th almost period follows from the $n$th best rational approximation $p_n/q_n$ to the irrational number $\beta$:
\begin{equation}
    T_n = \frac{2 \pi q_n}{\Omega_1}.
\end{equation}
The best rational approximations $p_n/q_n$ are obtained by truncating the infinite continued fraction expansion of $\beta$:
\begin{align}
   \beta &= a_0 + \frac{1}{a_1 + \frac{1}{a_2 + \frac{1}{a_3 + \ldots}}} \label{eq:cfrac}\\
\frac{p_n}{q_n} &= a_0 + \frac{1}{a_1 + \frac{1}{\dots \frac{1}{a_{n}}}}
\label{eq:pfrac}
 \end{align}
For example, if $\beta = (1+\sqrt{5})/2$, the golden ratio, then $a_n = 1$ and the best rational approximations are set by $p_{n-1} = q_n = F_{n}$ where $F_n$ is the $n$th Fibonacci number.

\subsubsection{A finite time notion of quasi-periodicity}

We cannot experimentally measure a frequency with infinite precision, and thus do not expect physical properties to depend on whether a quantity is rational or irrational. In this sub-section we give a precise condition for dynamics to be quasi-periodic in an experiment with a finite run time $t_\mathrm{exp}$.

The key property we need to preserve is that $\Bt_t$ is uniformly distributed over the torus. In the limit $t_\mathrm{exp} \to \infty$, this requirement is precisely defined as follows: if one chooses a random $t \in [0,t_\mathrm{exp}]$ with uniform probability, and an arbitrary convex region $A \subset (0,2\pi]^2$, then $\Bt_t \in A$ with probability $\mathrm{Prob}(\Bt_t \in A) = |A|/4 \pi^2$. $\Bt_t$ is uniformly distributed over the torus as $t_\mathrm{exp} \to \infty$ if and only if $\beta$ is irrational (Eq.~\eqref{eq:not_rat}).

We now consider finite $t_\mathrm{exp}$. Our aim is to choose $\beta$ such that for $t \in [0, t_\mathrm{exp}]$, $\Bt_t$ is approximately uniformly distributed. To make this notion precise, it suffices to consider $\Bt_t$ at the stroboscopic times $t_q = 2 \pi q /\Omega_1$ where $q$ is a positive integer such that $q \leq q_\mathrm{exp} := \lfloor t_\mathrm{exp} \Omega_1/ (2 \pi) \rfloor$. At these times, $\theta_2$ takes on values
\begin{equation}
    \theta_{t_q 2} = 2 \pi ( q \beta \mod 1 )
\end{equation}
The minimum separation of two values of $\theta_{t_q 2}$ is given by
\begin{equation}
    \Delta \theta_2 = \min_{p,q \leq \, q_\mathrm{exp}} 2 \pi q \left| \beta - \frac{p}{q}\right|
\end{equation}
where the $\min$ is taken over all positive integers $p,q$ such that $q \leq q_\mathrm{exp}$. This minimum spacing cannot be larger than that found for evenly spaced points
\begin{equation}
    \Delta \theta_2^\star = \frac{2 \pi}{q_\mathrm{exp}}
\end{equation}
The points $\theta_{t_q 2}$ are approximately uniformly distributed if the ratio of $\Delta \theta_2$ and $\Delta \theta_2^\star$ is close to unity. That is
\begin{equation}
    c : = \frac{\Delta \theta_2}{\Delta \theta_2^\star} = \min_{p,q \leq \, q_\mathrm{exp}} q \, q_\mathrm{exp} \left|\beta - \frac{p}{q}\right| \approx 1.
\end{equation}
We note that the minimand is minimised for $p/q$ equal to a best possible rational approximation (Eq.~\ref{eq:pfrac}). Thus
\begin{equation}
    c = q_n \, q_\mathrm{exp} \left|\beta - \frac{p_n}{q_n}\right|
\end{equation}
where $p_n/q_n$ is the best rational approximation to $\beta$ with the largest denominator such that $q_n \leq q_\mathrm{exp}$ holds.

If $\beta$ is rational such that $\Bt_t$ starts to repeat before $t < t_\mathrm{exp}$ then $c = 0$. If $\beta$ is sufficently close to a low order rational that the drives appear periodic on experimental timescales, then one finds $0 < c \ll 1$. Conversely if $c$ is close to unity, then $\Bt_t$ is roughly evenly spread over the torus up to the time $t = t_\mathrm{exp}$.

We note that requiring $c$ to have a value close to unity does not require fine tuning, and is typically true for $\beta$ drawn from the uniform measure. We further note that by Hurwitz's theorem, in the limit of large $q_\mathrm{exp}$, $c$ is upper bounded: $c \leq 1/\sqrt{5}$. This bound is saturated for $\beta$ equal to the golden ratio, the ``most'' irrational number. We therefore used the golden ratio as the value of $\beta$ for the experimental measurements presented in the manuscript.

\subsection{Quasi-energy states}
\label{app.QES}

Consider the standard time evolution operator 
\begin{equation}
    U(t;\Bt_0) := \mathscr{T} \exp \left( - i \int_0^t \d t H(\Bt_0 + \Bw t) \right).
\end{equation}
In periodically driven systems, the Floquet quasi-energy states are the eigenvectors of the time evolution operator after an integer number of periods. Analogous quasi-energy states can be defined for the two-tone quasi-periodically driven system as the limiting value of the eigenvectors of the time evolution operators at almost periods. Define:
\begin{equation}
    U(T_n;\Bt_0) \ket{\phi_n^j(\Bt_0)} = \e^{- i T_n \epsilon_n^j(\Bt_0)} \ket{\phi_n^j(\Bt_0)}
\end{equation}
where $j$ enumerates the different eigenstates. For a two level system, $j=1,2$. For smooth $H(\vec{\theta})$, Refs.~\cite{Martin2017,Crowley2019} showed the eigenstate projectors converge as $n \to \infty$ at all $\Bt_0$:
\begin{equation}
    \ket{\phi^j(\Bt_0)}\bra{\phi^j(\Bt_0)} = \lim_{n \to \infty} \ket{\phi_n^j(\Bt_0)}\bra{\phi_n^j(\Bt_0)}. \label{Eq:QEstateQP}
\end{equation}
As the phase of $\ket{\phi_n^j(\Bt_0)}$ is not defined, the quasi-energy states themselves do not converge. For convenience, we work in a gauge in which the $\ket{\phi^j(\Bt_0)}$ are locally smooth functions of $\Bt_0$ and the derivative $\nabla_{\Bt_0} \ket{\phi^j(\Bt_0)}$ exists. Our conclusions will of course be gauge independent.

\subsubsection{Quasi-energy states as a basis for time evolution}

As with Hamiltonian eigenstates in the static case and Floquet states in the periodic case, the quasi-energy states defined in Eq.~\eqref{Eq:QEstateQP} are solutions to the time-dependent Schr\"odinger equation when multiplied by a phase factor. That is, 
\begin{equation}
    \ket{\psi_t} = \e^{- i \alpha_j(t;\Bt_0)} \ket{\phi^j(\Bt_t)}
\end{equation}
is a solution to $i\partial_t \ket{\psi_t} = H(\Bt_t) \ket{\psi_t}$ with phases 
\begin{equation}
\begin{aligned}
    \alpha_j(t;\Bt_0) &= \int_0^t \d s \left( E_j(\Bt_s) - \Bw \cdot \BA_j(\Bt_s) \right) 
    \\
    &= \int_0^t \d s E_j(\Bt_s) - \int_{\Bt_0}^{\Bt_t} \d \Bt_s \cdot \BA_j (\Bt_s)
\end{aligned}
\label{eq:alpha}
\end{equation}
Above
\begin{align}
    E_j(\Bt_t) &= \bra{\phi^j(\Bt_t)} H(\Bt_t) \ket{\phi^j(\Bt_t)}
    \\
    \BA_j(\Bt) &= i \bra{\phi^j(\Bt_t)} \nabla_{\Bt} \ket{\phi^j(\Bt_t)}
\end{align}
are the instantaneous energy and the (gauge dependent) Berry connection respectively, and the line integral in Eq.~\eqref{eq:alpha} is taken along the path $\Bt_s$ for $s \in [0,t]$. The two terms in Eq.~\eqref{eq:alpha} correspond to the usual dynamical and geometric phases respectively.

A general solution to the time-dependent Schr\"odinger equation $\ket{\psi_t}$ thus decomposes as:
\begin{equation}
    \ket{\psi_t}  = \sum_j \ket{\phi^j(\Bt_t)} \e^{- i \alpha_j(t; \Bt_0)} \braket{\phi^j(\Bt_0)}{\psi_0}.
\end{equation}

\subsection{Topological classification}
\label{app.topoc}

As is familiar for quantum states defined over the torus $\mathbb{T}^2$, the quasi-energy states may be classified according to an integer-valued topological index, the Chern number $C_j$, given by the integrated Berry curvature
\begin{equation}
     C_j = \frac{1}{2 \pi} \int_{\mathbb{T}^2} d^2 \vec{\theta} \, \mathscr{B}_j (\vec{\theta}) \, \in  \mathbb{Z} \label{Eq:DefQEChern}
\end{equation}
where, as usual, the (gauge invariant) Berry curvature, is given by
\begin{equation}
    \mathscr{B}_j = 2 \operatorname{Im} \braket{\partial_{\theta_2}\phi^j}{\partial_{\theta_1}\phi^j}.
\end{equation}
In the remainder of this section, we show how the Chern number determines the oscillation frequency of the overlap/fidelity and other observables.

\subsection{Dynamical signature: Topological oscillations}

In this section, we compute the phase difference between two trajectories that start at slightly different initial drive phases. The phase difference is set by the integrated Berry curvature of the region enclosed by the two trajectories in $\Bt$-space. This phase difference determines the difference between expectation values at the end points of the trajectories. 

The expectation value of a Hermitian operator $O$ at time $t$ is:
\begin{equation}
    O(t;\Bt_0) = \qexp{U(t;\Bt_0)^\dag O U(t;\Bt_0)}{\psi_0}.
\end{equation}
Decomposing in the quasi-energy basis:
\begin{equation}
    O(t;\Bt_0) = \sum_{ij} O_{ij}(t;\Bt_0)
    \label{eq:O_dec}
\end{equation}
where
\begin{equation}
\begin{aligned}
    O_{ij}(t;\Bt_0) = & \braket{\psi_0}{\phi^i(\Bt_0)}\bra{\phi^i(\Bt_t)}O\ket{\phi^j(\Bt_t)}\braket{\phi^j(\Bt_0)}{\psi_0}
    \\
    & \, \times \e^{i(\alpha_i(t;\Bt_0) - \alpha_j(t;\Bt_0))}.
\end{aligned}
\end{equation}

Suppose the initial drive phase $\Bt_0$ is perturbed by a small amount $\dBt$. We will show below that
\begin{align}
    O_{ij}(t;\Bt_0 + \dBt) = &
    O_{ij}(t;\Bt_0) \e^{i (C_i - C_j) \omega_{\mathrm{T}} t} + \mathrm{O}(|\dBt| t^0),
    \label{Eq:TopOscDef}
\end{align}
where
\begin{align}
    \omega_{\mathrm{T}} &\equiv \frac{|\dBt \times \Bw|}{2 \pi} \label{Eq:TopFreqDef},
\end{align}
and the Chern number $C_{j}$ is defined in Eq.~\eqref{Eq:DefQEChern}. The phase factor that appears in the first term in the RHS of Eq.~\eqref{Eq:TopOscDef} oscillates with a frequency set by the difference in Chern numbers, and thus is topological in origin. This frequency appears in the difference of the expectation value of $O$ between trajectories with perturbed initial phases:
\begin{equation}
\begin{aligned}
    \Delta O(t;\Bt_0)^2 &: = \left[ O(t;\Bt_0 + \dBt/2) - O(t;\Bt_0-\dBt/2) \right]^2 
    \\
    & = 16 \left[ \sum_{i<j} \sin \left( \tfrac12 (C_i - C_j) \omega_{\mathrm{T}} t \right) \operatorname{Im} O_{ij}(t;\Bt_0) \right]^2 \\
    &+ \mathrm{O}(|\dBt| t^0) \label{Eq:GenOSq}
    \end{aligned}
\end{equation}
For the case of a two level system $C = C_1 = -C_2$, Eq.~\eqref{Eq:GenOSq} simplifies 
\begin{align}
    \!\!\Delta O(t;\Bt_0)^2 &= 8 \left( 1 - \cos ( 2 C \omega_\mathrm{T} t ) \right) \left(\mathrm{Im} O_{12}(t;\Bt_0) \right)^2 \nonumber \\
    &+ \mathrm{O}(|\dBt| t^0).
        \label{eq:DeltaO}
\end{align}
The off-diagonal matrix element $O_{12}(t;\Bt_0)$ varies at the drive frequencies $\Omega_1, \Omega_2$, while the front-factor oscillates with the `topological' frequency $2 C \omega_\mathrm{T} $. As $|\dBt| \to 0$, $ \omega_\mathrm{T} \ll \Omega_{1,2}$, so that $\Delta O(t;\Bt_0)^2$ is bounded by an oscillatory envelope with frequency set by the Chern number $C$. This envelope is visible in the experimental data taken in the topological regime of the dynamics (Fig.~\ref{fig.divergence.diff}(c)) and is used to experimentally extract the Chern number.

Note that $\Delta O(t;\Bt_0)^2$ is bounded by an oscillatory envelope for general few level systems, so that this result is not specific to the two level case. This is visible in Eq.~\eqref{Eq:GenOSq} where, due to the sinusoidal factor on each term in the sum, $\Delta O(t;\Bt_0)^2 = \mathrm{O}(|\dBt| t^0)$ small at times $\omega_\mathrm{T} t \in 2 \pi \mathbb{N}$ owing to the fact that $C_i,C_j \in \mathbb{Z}$.

\subsubsection*{Proof of Eq.~\eqref{eq:DeltaO}}

To derive~\eqref{eq:DeltaO}, we consider the terms of~\eqref{eq:O_dec} in turn. The diagonal terms $O_{ii}(t;\Bt_0)$ are real, whereas the off-diagonal terms are constrained by Hermittivity $O_{ij}(t;\Bt_0) = O_{ji}(t;\Bt_0)^*$. The phase of $O_{ij}(t;\Bt_0)$ is set by
\begin{widetext}
\begin{equation}
\begin{aligned}
    \chi_{ij}(t;\Bt_0) := \arg \left[ O_{ij}(t;\Bt_0) \right] 
     = \alpha_i(t;\Bt_0) - \alpha_j(t;\Bt_0) + \arg\left[\braket{\psi_0}{\phi^i(\Bt_0)}\bra{\phi^i(\Bt_t)}O\ket{\phi^j(\Bt_t)}\braket{\phi^j(\Bt_0)}{\psi_0}\right]
\end{aligned}
\end{equation}
where $\arg(r \e^{i \chi}) = \chi$ is the phase of a complex number. For $i \neq j$, the phase $\chi_{ij}(t;\Bt_0)$ is sensitive to the perturbation $\dBt$. As usual, the phase alone is not a physical quantity. However, the phase difference induced by the perturbation is physically accessible. Thus, we consider

\begin{equation}
\begin{aligned}
    \delta \chi_{ij}(t;\Bt_0) :=& \chi_{ij}(t;\Bt_0 + \dBt) - \chi_{ij}(t;\Bt_0)
    \\
     = & \int_0^t \d s \left( E_i(\Bt_s + \dBt) - E_i(\Bt_s) - E_j(\Bt_s + \dBt) + E_j(\Bt_s)\right) + \oint_{\partial A} \d \Bt' \cdot \left(\BA_i (\Bt') - \BA_j (\Bt') \right) + \mathrm{O}(|\dBt| t^0)
     \label{eq:aaa}
\end{aligned}
\end{equation}
where the second line follows from Eq.~\eqref{eq:alpha}, and standard algebraic manipulation. Let us unpack Eq.~\ref{eq:aaa}: in obtaining the equation the foreboding $\operatorname{arg}\left[ \dots \right]$ terms give rise only to the $\mathrm{O}(|\dBt| t^0)$ correction. The phase terms are written in terms of $\partial A$, which denotes the boundary of the region $A$ defined by $\Bt_0 + u \dBt + s \Bw$ for $u \in [0,1]$, $s \in [0,t]$. This region has area $|A| = t |\dBt \times \Bw|$, and is shown in Fig.~\ref{fig.seq} in the main text. Using Stokes theorem, the phase difference is re-written in terms of the Berry curvature
\begin{equation}
\begin{aligned}
    \delta \chi_{ij}(t;\Bt_0) &= \int_A \d^2 \Bt' \, \frac{\dBt}{|\dBt|} \cdot \nabla_{\Bt} \left( E_i (\Bt') - E_j (\Bt') \right) + \int_A \d^2 \Bt' \left(\mathscr{B}_i (\Bt') - \mathscr{B}_j (\Bt') \right) + \mathrm{O}(|\dBt| t^0).
    \label{eq:delta_chi_result}
\end{aligned}
\end{equation}
In the limit of long time and small $\dBt$, $A$ converges to a uniform sampling of the torus. At finite $t$ such that $\Omega_1 t,\Omega_2 t \gg 1$ the order of corrections to this is set by the spacing of trajectories $\Delta \theta \approx 2 \pi/|\Bw|t$
\begin{equation}
\begin{aligned}
    \int_A \d^2 \Bt' \mathscr{B}_i(\Bt') = \frac{t |\dBt \times \Bw|}{4 \pi^2} \left[ \int \d^2 \Bt'\,\mathscr{B}_i(\Bt') + \mathrm{O}\left( \frac{2 \pi}{|\Bw| t}\max_{\Bt} | \nabla_{\Bt} \mathscr{B}_i(\Bt) | \right) \right]
     = \frac{|\dBt \times \Bw| C_i t}{2 \pi} + \mathrm{O}(|\dBt| t^0)
\label{eq:Chern_numer}
\end{aligned}
\end{equation}
\end{widetext}
and by the same argument one finds
\begin{equation}
        \int_A \d^2 \Bt' \, \frac{\dBt}{|\dBt|} \cdot \nabla_{\Bt} E_i (\Bt') = \mathrm{O}(|\dBt| t^0).
\label{eq:grad_E}
\end{equation}
From the results~\eqref{eq:delta_chi_result},~\eqref{eq:Chern_numer} and~\eqref{eq:grad_E} it follows that
\begin{equation}
    \delta \chi_{ij}(t;\Bt_0)  = (C_i - C_j ) \omega_\mathrm{T} t  + \mathrm{O}(|\dBt| t^0)
    \label{eq:delta_chi}
\end{equation}
with $\omega_{\mathrm{T}}$ given by Eq.~\eqref{Eq:TopFreqDef}. Here we have analaysed the phase; the corresponding analysis of the modulus $r_{ij}(t;\Bt_0) = |O_{ij}(t;\Bt_0)|$ is significantly simpler and yields 
\begin{equation}
    r_{ij}(t;\Bt_0 + \dBt) = r_{ij}(t;\Bt_0) + \mathrm{O}(|\dBt| t^0)
    \label{eq:r}
\end{equation}
Combining Eqs.~\eqref{eq:delta_chi}, and~\eqref{eq:r} yields Eq.~\eqref{Eq:TopOscDef}, from which Eqs.~\eqref{Eq:GenOSq} and ~\eqref{eq:DeltaO} follow.

\subsection{State overlap oscillations}
\label{app.SOO}

A particularly stark application of the result derived in the previous section is to the state overlap between two trajectories following a perturbation. To relate the previous result to the overlap we consider a family of operators $O^\mu$ rather than the single operator $O$, and restrict to the case where the $O^\mu$ form an orthonormal operator basis, i.e. that the operator matrix elements $O^\mu_{ij}$ satisfy
\begin{equation}
    \sum_\mu  O^{\mu}_{ij} O^{\mu}_{mn} = \delta_{in} \delta_{jm}
    \label{eq:ortho_ops}
\end{equation}
A well know example of the above relation in the 2-dimensional case are the normalized Pauli matrices $O^{(\mu)} = \sigma^\mu/2$ for $\mu = 0,1,2,3$ (with $\sigma^0$ the identity). 

Eq.~\eqref{eq:ortho_ops} has two useful corollaries which we will use
\begin{align}
    \sum_\mu  \bra{\alpha} O^\mu \ket{\alpha}  \bra{\beta} O^\mu \ket{\beta} & = \braket{\alpha}{\beta}\braket{\beta}{\alpha},
    \label{eq:orthocorr1}
    \\
\sum_\mu \left( \mathrm{Im}  \bra{\alpha} O^\mu \ket{\beta} \right)^2 &= \tfrac12\braket{\alpha}{\alpha}\braket{\beta}{\beta}.
    \label{eq:orthocorrolary}
\end{align}
For such a complete orthonormal basis of operators, the results of the previous section may be related to the overlap $F = |\braket{\psi'}{\psi}|^2$ between the unperturbed state $\ket{\psi} = U(t;\Bt_0 - \dBt/2)\ket{\psi_0}$ and the perturbed state $\ket{\psi'} = U(t;\Bt_0 + \dBt/2)\ket{\psi_0}$ in the following way
\begin{equation}
\begin{aligned}
    \!\sum_\mu \Delta O^{\mu}(t;\Bt_0)^2 =& \sum_\mu \left( O^{\mu}(t;\Bt_0  + \dBt / 2) \right)^2 
    \\
    & \,
    + \sum_\mu \left( O^{\mu}(t;\Bt_0 - \dBt/2) \right)^2 \\
    & \, - 2\sum_\mu  O^{\mu}(t;\Bt_0 + \dBt/2)   O^{\mu}(t;\Bt_0 - \dBt / 2 ) 
    \\
    = & 2 - 2 \left| \qexp{U_t^\dag U_t'}{\psi_0} \right|^2
    \\
     = & 2 - 2 F.
    \label{eq:OtoFidelity}
\end{aligned}
\end{equation}
where we write $U_t = U(t;\Bt_0 + \dBt/2)$ and $U_t' = U(t;\Bt_0 - \dBt/2)$ for compactness.
Moreover, for a two level system, using the relations~\eqref{eq:DeltaO},~\eqref{eq:OtoFidelity} and~\eqref{eq:orthocorrolary} it follows that
\begin{equation}
\begin{aligned}
    \!F& =  1 - \tfrac12  \sum_\mu \left(\Delta O^{\mu}(t;\Bt_0) \right)^2 
    \\
    & = 1 - 4 \left( 1 - \cos (2 C\omega_\mathrm{T} t) \right) \sum_\mu \left(\mathrm{Im} O_{12}^\mu(t;\Bt_0) \right)^2 + \mathrm{O}(|\dBt|t^0)
    \\
    & = 1 - 2 \left( 1 - \cos (2 C\omega_\mathrm{T} t) \right) P_1 P_2  + \mathrm{O}(|\dBt|t^0)
    \label{eq:Fid2}
\end{aligned}
\end{equation}
where, as before, $C = C_1 = -C_2$, $\omega_\mathrm{T} = |\dBt \times \Bw|/(2 \pi)$, and  $P_j = |\braket{\psi_0}{\phi^j(\Bt_0)}|^2$. Thus for $P_1 = P_2 = \tfrac12$ we find the final result of this section
\begin{equation}
    F = \tfrac{1}{2} \left( 1 + \cos (2 C \omega_\mathrm{T} t) \right) + \mathrm{O}(|\dBt|t^0).
    \label{eq:Fid}
\end{equation}

Two comments are in order. First, we recover the form used in the main text in Eq.~\eqref{eq:Chern_def} if we define $\Phi_i$ to be the Berry curvature enclosed by the trajectory
\begin{equation}
    \Phi_i = \int_A \d^2 \Bt' \mathscr{B}_i(\Bt') 
\end{equation}
and consider the two band case where the two bands have equal and opposite Berry curvature $\mathscr{B}_1 = - \mathscr{B}_2$, as in the main text, so that
\begin{equation}
    \Phi = \Phi_1 - \Phi_2 =  2 C \omega_\mathrm{T} t + \mathrm{O}(|\dBt| t^0).
\end{equation}
Using Eq.~\eqref{eq:Fid2} we then obtain
\begin{equation}
    \begin{aligned}
    \label{eq.app.overlap}
    F &= 1 - 4 P_1 P_2 \sin^2 (\Phi/2)
    \\
    & = 1 - 4 P_1 P_2 + 4 P_1 P_2 \cos \Phi
    \end{aligned}
\end{equation}
hence for $P_1=P_2= \tfrac12$ we recover $F = \cos^2 (\Phi/2)$ (Eq.~\eqref{eq:Chern_def}). 

Finally, we note that for small $\omega_\mathrm{T} t$ we recover the linear divergence of trajectories noted in Ref.~\cite{Crowley2019}. Specifically, if we take the Bures distance $D_\mathrm{B}(\psi',\psi) = \arccos |\braket{\psi'}{\psi}|$ as a metric over quantum states, then, using Eq.~\eqref{eq:Fid}, we find this distance exhibits initially linear growth
\begin{equation}
\begin{aligned}
    D_\mathrm{B}(\psi',\psi) &= 2 C\omega_\mathrm{T} t\sqrt{P_1 P_2} + \mathrm{O}( C\omega_\mathrm{T} t)^3 + \mathrm{O}(|\dBt|t^0)
    \\
    & = t \frac{|\dBt \times \Bw| \sigma(C)}{2 \pi} + \mathrm{O}(C \omega_\mathrm{T} t)^3 + \mathrm{O}(|\dBt|t^0)
\end{aligned}
\end{equation}
The linear growth rate is in agreement with Eq. (52) of Ref.~\cite{Crowley2019}, where we have used
\begin{equation}
\sigma^2(C) = \sum_{j=1}^2 C_j^2 P_j - \bigg( \sum_{j=1}^2 C_j P_j \bigg)^2.
\end{equation}

\section{Implementation of the Counterdiabatic Potential}
\label{app.VCD}

\begin{figure*}
	\centering
	\includegraphics[width= 1\textwidth]{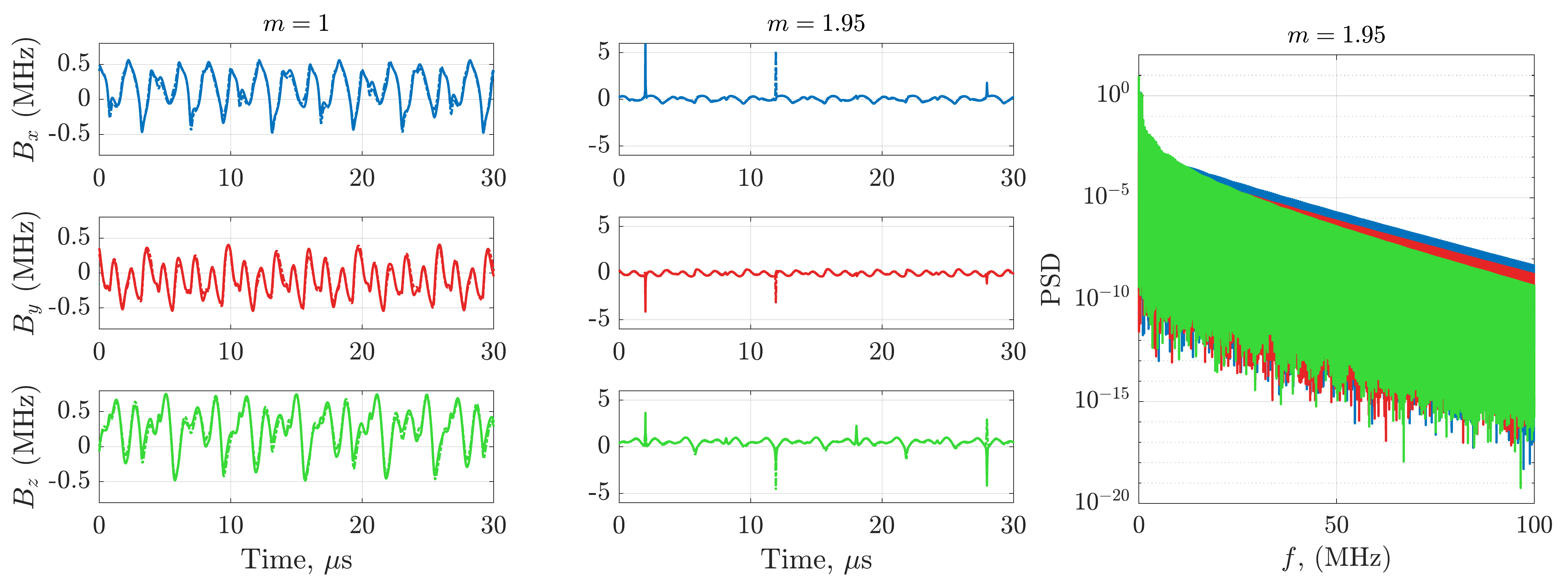}
	\caption{\textbf{Left, middle:} Time dependence of the magnetic field components of $H_{\mathrm{CD}}$ in the topological regime and near the critical point. Note the large `spikes' that occur near the critical point due to $V_{\mathrm{CD}}$. Hamiltonian parameters: $\gamma B_0 = 2\pi \times 0.25 \uu{MHz}, \, m = 1, \, 1.95,  \, \vec{\Omega}=2\pi \left(0.5, 0.5\varphi\right) \uu{MHz}, \, \Bt_0 = \left(0, \pi/2 \right), \, \dBt_0 = \left(\varphi/8, -1/8 \right)$. \textbf{Right:} Exponentially decaying power spectral density of the components of $\vec{B}_\mathrm{CD}$ near the critical point, $m = 1.95$}
	\label{fig.app.cdfields}
\end{figure*}

The topological dynamics persists as long as the system remains in the lower band\cite{Crowley2019, Martin2017}. If the system excites due to non-adiabatic effects, the dynamics appears trivial. Such excitation can be suppressed by driving the system adiabatically slowly. However this is experimentally infeasible as driving slowly lengthens the period of the topological oscillations such that it becomes comparable to experimental timescale, which is limited by decoherence. Instead we use a counterdiabatic potential~\cite{demirplak2003adiabatic,berry2009transitionless,del2013shortcuts,sels2017minimizing} to prevent the qubit from exciting even when driving at finite rate. The counterdiabatic potential $V_{\mathrm{CD}}$ suppresses transitions between the instantaneous eigenstates. In general this term satisfies equation \cite{sels2017minimizing}:
\begin{equation}
    \label{eq:CD_general}
    [ i \partial_{t}H + [H,V_{\mathrm{CD}}], H ] = 0.
\end{equation}
For a two-level system with $\tr{H}=0$, a solution is:
\begin{equation}
    \label{eq:VCD_qubit}
    V_{\mathrm{CD}} = \frac{i}{2}\frac{[\partial_t H, H]}{\tr {H^2}}.
\end{equation}
For the Chern insulator Hamiltonian (Eq. \ref{eq:CI_t}) one finds
\begin{equation}
    V_{\mathrm{CD}}(t) = \vec{B}_{\mathrm{CD}}(\Bt_t) \cdot \vec{\sigma}
\label{eq:VCD_fullForm}
\end{equation}
where $\vec{B}_{\mathrm{CD}}$ has elements
\begin{align*}
    \!\!\! B_{\mathrm{CD}}^x &= -\frac{\Omega_2 - m \Omega_2 \cos\theta_{2} + \Omega_2 \cos\theta_{1} \cos\theta_{2} + \Omega_1 \sin\theta_{1} \sin\theta_{2} }{4 + 2m^2 - 4 m \cos \theta_{1} - 4 m \cos \theta_{2} + 4 \cos\theta_{1} \cos\theta_{2}}, \\
    \!\!\!B_{\mathrm{CD}}^y &= \frac{\Omega_1 - m \Omega_1 \cos\theta_{1} + \Omega_1 \cos\theta_{1} \cos\theta_{2} + \Omega_2 \sin\theta_{1} \sin\theta_{2} }{4 + 2m^2 - 4 m \cos \theta_{1} - 4 m \cos \theta_{2} + 4 \cos\theta_{1} \cos\theta_{2}}, \\
    \!\!\!B_{\mathrm{CD}}^z &= \frac{\Omega_2 \cos\theta_{2} \sin\theta_{1} - \Omega_1 \cos\theta_{1} \sin\theta_{2} }{4 + 2m^2 - 4 m \cos \theta_{1} - 4 m \cos \theta_{2} + 4 \cos\theta_{1} \cos\theta_{2}}. \\
\end{align*}

Away from the critical points $|m|=0,2$, $B_{\mathrm{CD}}$ is smooth, and thus asymptotically the Fourier coefficients are exponentially decaying in their order. Specifically, for $m = 1,3$ we find that the amplitudes of the Fourier components have decreased by a factor greater than $10^{10}$ at the frequency $\omega = 40\Omega_1$. For typical experiments where $\Omega_1 = 2\pi \times 0.5 \text{MHz}$, this translates to a bandwidth requirement of only $20 \text{MHz}$, significantly less than the $100 \text{MHz}$ bandwidth of our arbitrary wavefrom generator (AWG). Even at $m = 1.95$, Fourier components at the $100 \text{MHz}$ maximum frequency are reduced by a factor of $ 10^{7} $ compared to that at zero frequency, see Fig.~\ref{fig.app.cdfields}. Thus, the counterdiabatic potential can be implemented on our hardware.

At the critical points $V_{\mathrm{CD}}$ diverges at the band touching points (for $|m|=2$ this occurs at $\vec{\theta} = (0,0)$). When $\vec{\theta}_t = \vec{\theta}_0 + \vec{\omega} t$ comes close to these points, the amplitude of the terms of $\vec{B}_{\mathrm{CD}}$ becomes correspondingly large and can lead to clipping of the waveform if it exceeds the maximum supported by the amplifiers. This causes the counterdiabatic potential to be imperfectly implemented and leads to excitations between the instantaneous bands of the Hamiltonian, contributing to the decay of the overlap oscillations. 

\section{Experimental setup and measurement details}

\subsection{Experimental setup}
\label{app.experimentSetup}

In this appendix, we describe the experimental setup used in this work. We note that this setup is similar to that used in a previous related study by our group and so the description is very similar, see \cite{boyers2019floquet}.

\begin{figure}
	\centering
    \includegraphics[scale=.6]{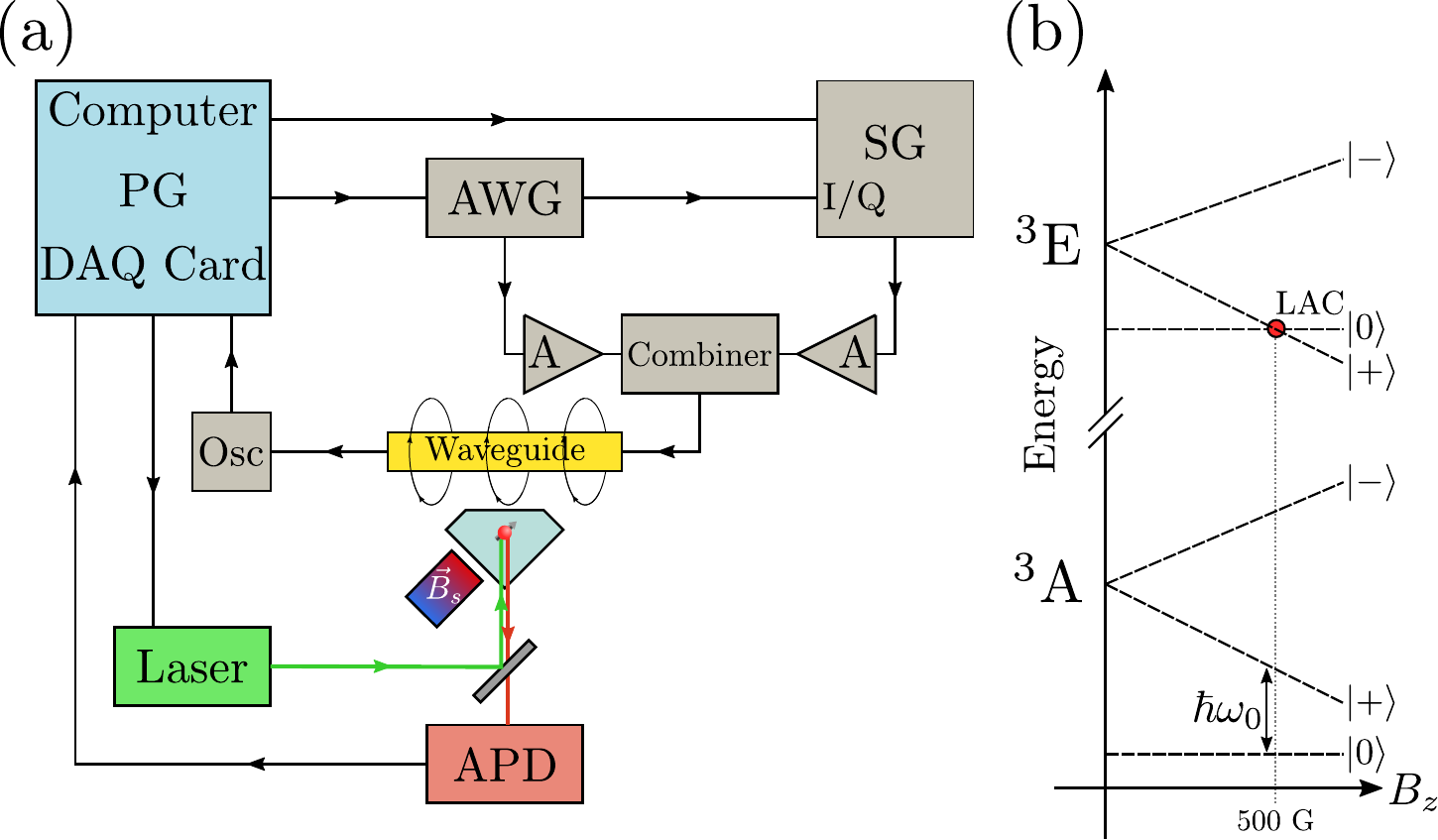}

	\caption{ \textbf{(a)} Schematic diagram of hardware setup. PG: Pulse Generator; DAQ Card: Data Acquisition card; AWG: Arbitrary Waveform Generator; SG: Signal Generator; A: Amplifier; Osc: Oscilloscope; APD: Avalanche Photodiode. Laser module includes a double pass acoustic-optic modulator (AOM).
	\textbf{(b)} Energy levels of the NV center under a static magnetic field along the NV symmetry axis, taken to be the z-axis. $^3$A and $^3$E refer to the orbital energy levels while kets refer to the electronic spin energy levels.}
	\label{fig.app.hardware}
\end{figure}

The diamond used in our experiments was grown by $\text{C}^{12}$ enriched carbon vapor deposition and bombarded with $\text{N}^{15}$ ions and annealed to produce NV centers. In Figure \ref{fig.app.hardware} panel a) we show a schematic diagram of the experimental apparatus used to probe and manipulate individual NV centers. The setup is operated by a computer which controls the hardware and has a pulse generator card (PG) for creating trigger pulses and a data acquisition card (DAQ card) for receiving single photon counts from the avalanche photodiode (APD). We address individual NV centers using a 532nm laser in a home-built scanning confocal microscope setup which uses the APD to detect fluorescence. An acoustic-optic modulator (AOM) allows us to create laser pulses with a minimum duration of 100ns. A bar magnet ($\vec{B}_s$) mounted on a 5 axis translation/rotation stage is aligned with the NV center symmetry axis, taken to be the z-axis, and the distance from the NV center is tuned to produce the desired static field.

To create an effective two level system from the NV center, we tune the static field $\vec{B}_s$ to the NV center excited state level anti-crossing (LAC) at approximately 500G, as shown in the energy level diagram in Figure \ref{fig.app.hardware} panel b). At the LAC, optically pumping the NV center simultaneously polarizes both the NV spin and the nuclear spin \cite{Jacques2009}. Since the nuclear spin has a much longer relaxation time and much smaller gyromagnetic ratio than the electronic spin, the nuclear spin very nearly remains in the ground state throughout each measurement, and it merely shifts the NV electronic spin transition frequency. Additionally, at the LAC, the NV spin states $\ket{+1}$ and $\ket{-1}$ are split by ${\sim}3$ GHz, allowing us to drive on resonance with the $\ket{0} \leftrightarrow \ket{+1}$ transition at $\omega_0 \approx 1.46$ GHz while only driving a negligible population into the $\ket{-1}$ state. The NV spin states $\ket{0}$ and $\ket{+1}$ serve as the two levels of an effective qubit, which we label as $\ket{\uparrow_z}$ and $\ket{\downarrow_z}$, respectively.

We manipulate the qubit using time-dependent magnetic fields $B_{x,y,z}$ generated by current in a waveguide placed near the NV center. To create $B_x(t)$ and $B_y(t)$, we generate voltage signals with the desired time dependence using an arbitrary waveform generator (AWG). We use them to perform I/Q modulation of a radio frequency carrier signal at frequency $\omega_0$ created by the signal generator (SG) so that $B_x(t)$ and $B_y(t)$ are encoded as the amplitudes of the in-phase and quadrature components of the carrier. $B_z(t)$ is also generated by an AWG, but is not modulated. Both the radio frequency and $B_z(t)$ signals are then amplified, combined, and delivered to a waveguide where they create a magnetic field at the NV center. The magnetic field will have a component parallel to the NV z-axis and a component perpendicular to it, which we take to be the lab-frame x-axis. The field generated by the $B_z(t)$ signal has frequency components up to at most $100 \, \text{MHz} \ll \omega_0 \sim 1 \text{GHz}$, so it cannot drive transitions and has negligible effect on the x-axis. Conversely, since $B_x(t)$ and $B_y(t)$ are modulated at $\omega_0$, much faster than any other scale in the system, in the rotating frame the z-axis field they contribute rapidly averages to zero and their effect on that axis can be ignored. So, the experimentally accessible Hamiltonian for the effective qubit is:
\begin{equation}
H_{\text{lab}} = \left(\frac{\omega_0}2 + B_z(t)\right) \sigma_z + 2(B_x(t)\cos{\omega_0t} + B_y(t)\sin{\omega_0t}) \sigma_x
\end{equation}

Since the drive amplitudes, $B_x(t)$ and $B_y(t)$, and the detuning, $B_z(t)$, are much smaller than the carrier frequency, we transform to the rotating frame with frequency $\omega_0$ and invoke the rotating wave approximation to give the following Hamiltonian:
\begin{equation}
H_{\text{rot}} = B_z(t) \sigma_z + B_x(t) \sigma_x + B_y(t) \sigma_y
\end{equation}
This allows us to implement arbitrary time dependent Hamiltonians by choosing $B_{x,y,z}$ appropriately, so long as the frequency components and amplitudes are much smaller than the carrier frequency $\omega_0 \approx 1.46 \text{GHz}$.

To calibrate the amplitudes of the fields $B_{x,y}$, we set them to be constant to drive Rabi oscillations and tune the power of the signal generator to give the desired Rabi frequency. To calibrate $B_{z}$, we set it to be constant and perform electron spin resonance (ESR) to measure how much the transition frequency shifted. The amplitude of the AWG voltage signal is then adjusted until the frequency shift corresponds to the desired value of $B_{z}$.

The Hamiltonians in the main text can then be implemented by programming the arbitrary waveform generators. To mitigate the effect of small detunings arising from parameter drift, we rotate the model~\ref{eq:CI_t} by $-\pi/2$ about the y-axis before implementing it in the NV center so that the $x$-axis of the NV center rotating frame is the $z$-axis of the model, the NV $z$-axis is the model $x$-axis, and the $y$-axis is the same. This prevents small DC offsets which occur on the NV $z$-axis from changing the value of $m$ in the model. We have:
\begin{align}
    \label{eq:CI_VR}
    B_x(t) =&  B_0 \left( m - \cos \theta_1 - \cos \theta_2 \right) - B_{CD,z}(t)  \\
    B_y(t) =& B_0  \sin \theta_2 + B_{CD,y}(t)  \\
    B_z(t) =& B_0 \sin \theta_1 + B_{CD,x}(t)
    \label{eq:CI_VR3}
\end{align}
Where $\theta_{1} = \Omega_1 t + \theta_{01}$, $\theta_{2} = \Omega_2 t + \theta_{02}$.

\subsection{Measurement details and Data analysis}
\label{app.measurement_detail}
\begin{figure}[ht]
	\centering
    \includegraphics[scale=.8]{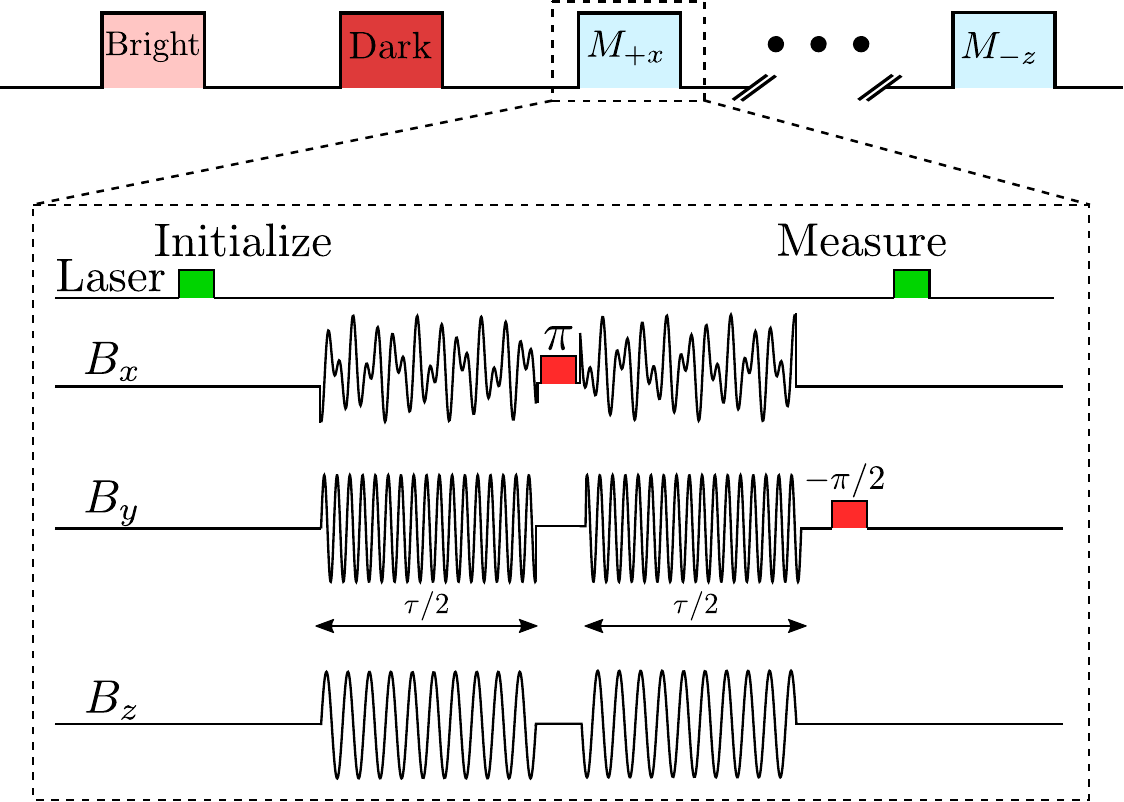}
    \caption{Pulse sequence for experiment measuring overlap. In the sections labeled `Bright' and `Dark', the fluorescence intensity corresponding to the states  $\ket{m_s = 0}$, $\ket{m_s = +1}$ is measured to normalize the measured projections, $M_{\pm \alpha}$. Then, each projection is measured. A laser pulse initializes the spin and it then evolves under the half-BHZ Hamiltonian (note that here $V_{\mathrm{CD}}$ is not included in the magnetic fields), followed by a $\pi$ spin-echo pulse and evolution under the half-BHZ Hamiltonian again. Finally, a $\pi/2$ pulse is applied and the projection along the measurement basis is measured. Here, the measurement of $M_{+x}$ is performed so a $-(\pi/2)_y$ pulse is used. Measurements of the other projections using different final pulses are represented by the ellipses.}
	\label{fig.app.seq}
\end{figure}

In this appendix, we provide additional details on the measurement procedure and data analysis. We begin with the pulse sequence used to implement the measurements, shown in Figure~\ref{fig.app.seq}. To suppress low frequency noise, we interleave evolution under the half-BHZ Hamiltonian with a spin-echo sequence. First, the NV is optically pumped with a laser pulse to initialize it in the $\ket{\uparrow_z}$ state, which is an even superposition of the initial quasi-energy eigenstates for any initial phase $\vec{\theta}_0 = \left( n \pi, \theta_{0,2}  \right)$, $n \in \mathbb{Z}$. Next, the spin evolves under the half-BHZ Hamiltonian for time $\tau/2$, followed by a $\pi_x$ pulse to flip the spin. The system then evolves under the half-BHZ Hamiltonian again for $\tau/2$ but with $B_y \rightarrow -B_y, \, B_z \rightarrow -B_z$ so that in the toggling frame that also rotates under the $\pi_x$ pulse, the orientation of the fields is the same before and after the pulse. Finally, a $-(\pi/2)_y$ pulse projects the $+x$ component of the spin onto the $z$-axis measurement basis and the population is read out by another laser pulse. In a full sequence, this sub-sequence is repeated with different $\pi/2$ pulses after the evolution in order to measure the six populations $M_{\pm \alpha} \equiv \langle P_{\pm \alpha} \rangle$, where
\begin{equation}
    P_{\pm \alpha} = \frac12\left( \mathbf{1} \pm \sigma_\alpha \right)
\end{equation}
for $\alpha = x,y,z$ measure the populations of the eigenvectors of each Pauli matrix. This procedure is repeated for each of the two initial phases $\vec{\theta}_0$, $\vec{\theta}_0+\dBt_0 $.

The setup does not directly measure $M_{\pm \alpha}$, but instead fluorescence counts. In a typical experiment, the pulse sequence is repeated $N_{\text{samp}} \sim 10^6$ times back-to-back for a given evolution time $\tau$ and the total counts $\widetilde{M}_{\pm \alpha}(\tau)$, are recorded for the spin starting with initial drive phases $\Bt_0$ and $\Bt_0 + \dBt$. This is then repeated for each value of $\tau$ and together these measurements represent one `average'. The experiment may then be repeated for $N_{\text{avg}} \sim 1 \, - \, 20$ averages to complete a run. Labeling the averages with a subscript $i=1, \ldots, N_{\text{avg}}$ as a function of $\tau$, the raw data is the collection of counts $\widetilde{M}_{\pm \alpha, i}(\tau)$

The physical measurement of fluorescence counts, $\widetilde{M}_{\pm \alpha, i}(\tau)$, will be a linear function of $M_{\pm \alpha}(\tau)$ due to collection efficiency, laser power, background counts, etc. To normalize $\widetilde{M}_{\pm \alpha, i}(\tau)$ we measure the `bright' counts by polarizing the spin and measuring the counts to give $\widetilde{b}_{i}(\tau)$, and the `dark' counts by polarizing the spin, applying a $\pi$ pulse, and then measuring to give $\widetilde{d}_{i}(\tau)$. Measurements of the bright and dark counts are done for each time point $\tau$ and average $i$. We normalize each average separately and then combine them to give $M_{\pm \alpha}(\tau)$:
\begin{equation}
\label{eq:normalization}
    M_{\pm \alpha}(\tau) = \frac{1}{N_{\text{avg}}} \sum_{i=1}^{N_{\text{avg}}} \frac{\widetilde{M}_{\pm \alpha, i}(\tau) - \widetilde{d}_{i}(\tau)}{\widetilde{b}_{i}(\tau) - \widetilde{d}_{i}(\tau)}
\end{equation}

To calculate the overlap from these measurements, we construct the expectation values as the difference of two measurements to subtract out common mode errors:
\begin{equation}
\label{eq:diff_expectation}
    \langle \sigma_\alpha \rangle = M_{+\alpha} - M_{-\alpha}
\end{equation}
Any pure state can be written in terms of Bloch sphere angles as:
\begin{equation}
\label{eq:psi_Bloch}
    \ket{\psi} = \cos \theta/2 \ket{\uparrow} +  e^{i \phi} \sin \theta/2 \ket{\downarrow}
\end{equation}
If we want the overlap, $F = \abs{\braket{\psi(t)}{\psi'(t)}}^2$ with another state $\ket{\psi'}$ (with primed variables) then we simply take a dot product, giving:
\begin{align}
    \begin{split}
        \abs{\braket{\psi(t)}{\psi'(t)}}^2 &= \tfrac12 \left(1 + \cos \theta \cos \theta' +  \sin \theta \cos \phi \sin \theta' \cos \phi' \right. \\
        &\left. + \sin \theta \sin \phi \sin \theta' \sin \phi' \right)
        \nonumber
    \end{split}
    \\
    &= \tfrac12\left(1 + \langle \sigma_z \rangle \langle \sigma_z \rangle' +  \langle \sigma_x \rangle \langle \sigma_x \rangle' + \langle \sigma_y \rangle \langle \sigma_y \rangle' \right)
    \label{eq:overPure}
\end{align}
This, combined with equation \ref{eq:diff_expectation} and \ref{eq:normalization} is how we construct the overlap presented in the main text. Note that since there is noise in the measurements $\langle \sigma_i \rangle$, there will be noise in $F$ that could cause it to be slightly above 1 or below 0. Although technically this is unphysical, it is clear that this is simply a result of noisy measurements and the interpretation is still clear: $F > 1$ corresponds to states that are very similar and $F<0$ corresponds to states that are nearly orthogonal. Since we use standard nonlinear regression to extract the Chern number from the overlap, the fact that $F$ can be above 1 or below 0 does not quantitatively impact our results in any significant way.

The overlap of two qubit states can also be defined for two density matrices, $\rho$, $\sigma$, relaxing the assumption that the states are pure:
\begin{align} 
    \nonumber
    F(\rho, \, \sigma) &= \left( \tr{\sqrt{\sqrt{\rho} \sigma \sqrt{\rho}}} \right)^2 \\
    \begin{split}
        &= \tfrac12 \left(1 + \langle \sigma_z \rangle \langle \sigma_z \rangle' +  \langle \sigma_x \rangle \langle \sigma_x \rangle' + \langle \sigma_y \rangle \langle \sigma_y \rangle' \right) \\
        & \quad +\tfrac12 \sqrt{(1-r^2)(1-r'^2)}
    \end{split}
\label{eq:fidelityRho}
\end{align}
This result is exactly that in \ref{eq:overPure} plus an extra term which accounts for the fact that unpolarized states are also similar and so have high overlap.

Since decoherence occurs during the experiment and the final state is thus a mixed state, it might seem reasonable to use this more general overlap rather than that for pure states. However, the overlap for mixed states obscures the effect of decoherence and makes the overlap a poor signature of the overlap oscillations. In the trivial regime, we expect that the overlap is nearly unity, $F \approx 1$. For two completely unpolarized states, the mixed state overlap of eq.~\ref{eq:fidelityRho} also gives $F = 1$ while the pure state overlap eq.~\ref{eq:overPure} gives $F = 0.5$. Thus, the mixed state overlap does not allow us to distinguish between states that are exhibiting the expected behavior and states that are completely unpolarized, while the pure state fidelity does. 

Similarly, in the topological regime, the pure state overlap undergoes oscillations about $F=\frac{1}{2}$ with decreasing amplitude, while the mixed state overlap will have more complicated behavior where the amplitude and offset of the oscillations change as the states decohere and the overlap approaches $F=1$. Thus, the pure state overlap more clearly shows the topological oscillations and can be fit with a model with fewer free parameters than the mixed state overlap. Since the oscillations are the phenomena we are ultimately interested in, we use the pure state overlap even as the states become mixed since it shows the oscillations more clearly.

We account for the decline in polarization by assuming that expectation values decay exponentially:
\begin{equation}
\label{eq:fidelityDecay}
    F = \tfrac12 + \tfrac12 \left(\langle \sigma_z \rangle \langle \sigma_z \rangle' +  \langle \sigma_x \rangle \langle \sigma_x \rangle' + \langle \sigma_y \rangle \langle \sigma_y \rangle' \right) e^{-t/\tau} 
\end{equation}
Where $\tau$ is the characteristic time scale for the decay of polarization. At $t \gg \tau$, the states have completely decohered and $F=0.5$. Combining this ansatz with the predicted form for the overlap of the states evolving under the half-BHZ Hamiltonian using equation \ref{eq.app.overlap} gives the expected behavior of the overlap with topological oscillations and decoherence:
\begin{equation}
\label{eq:fidelityExperPartial}
    F(t) = \tfrac12 + \left( \tfrac12 - 4 P_1 P_2 \sin^2\omega t    \right) e^{-t/\tau}
\end{equation}
Since $P_2 = 1 - P_1$, we can use a single parameter $A =  4 P_1 P_2$ when fitting the oscillations which we expect to be be near $1$ since we have chosen the initial states so that $P_1 \approx 1/2$. We also add a small offset parameter $\delta$ which ensures that the model can fit the mean of the data in the trivial regime where $F \approx 1$ but additive noise can make the mean slightly above or below 1. Together, these give the following functional form with fitting parameters $A, \, \omega_{\text{fit}}, \, \tau, \, \delta $:
\begin{equation}
    F(t) = \tfrac12 + \left( \tfrac12 -  A \sin^2 \omega_{\text{fit}}t     \right) e^{-t/\tau} + \delta
\end{equation}

We fit our data using non-linear least squares regression in MatLAB. Typical results give $\delta \sim \pm 0.01$ with a 95\% confidence interval generally consistent with $\delta = 0$, $A \approx 1$, and $\tau \sim 60-100 \, \mu \text{s}$. We extract the Chern number from the frequency by using the fact that the oscillation frequency is $\omega_{\mathrm{T}}C =  \frac{\left| \vec{\Omega} \times \dBt_0  \right|}{2 \pi}C$. Note that the form of the overlap means that we can not distinguish between positive and negative geometric phases so we report the absolute value of the ground state Chern number.

In the trivial phase there are no oscillations to fit to, only a small decline in the overlap. While this decline is likely a result of decoherence causing a loss of polarization, we put an upper bound on the Chern number in the trivial regime by assuming that there is no decoherence so that all the decline in the overlap is due to the geometric phase:
\begin{equation}
    F(t) = 1 - \sin^2 \omega_{\text{fit}}t + \delta
\end{equation}
We have also set $A=1$. When the argument of $A \sin^2\omega_{\text{fit}}t$ is small, it is easy to see by Taylor expansion that $A$ and $\omega_{\text{fit}}$ are redundant parameters. This causes the fitting algorithm to perform poorly because equally good fits can be achieved with a wide range of either parameter, resulting in spurious fits and large confidence intervals

\subsection{Error Analysis}
The trigger source in the pulse generator card and the trigger detector in the AWG both exhibit jitter, which lead to a measured maximum jitter of $3 \text{ns}$ as measured by the leading edge of a pulse from the AWG relative to a reference pulse from the PG card. This jitter leads to a small effective spread in the value of the phase perturbation, $\dBt_0$, which, in turn, leads to a small uncertainty in the measured Chern number. The measured Chern number is:
\begin{equation}
    \left| C_\textrm{exp} \right| = \frac{\omega_{\text{fit}}}{\omega_{\mathrm{T}}}
\end{equation}
Where $ \omega_{\mathrm{T}} = \abs{ \vec{\Omega} \times \dBt}/2 \pi$. Jitter thus causes an uncertainty in $\omega_{\mathrm{T}}$ since $\dBt$ has some spread. We propagate the error by computing the variance of $ \omega_{\mathrm{T}} $ while assuming that $\dBt_{\mathrm{tot}} = \dBt_{\mathrm{set}} + \left( \Omega_1 t_{j1}, \Omega_2 t_{j2} \right)$, where $\dBt_{\mathrm{set}}$ is the phase difference set experimentally and the second term is the phase difference due to jitter, with the jitter times $t_{j1}, t_{j2}$ being distributed uniformly over $\left[ -T, T \right]$. We find that:
\begin{equation}
    \sigma_{\omega_{\mathrm{T}}} = \sqrt{\tfrac23} \cdot \frac{\Omega_1 \Omega_2 T}{2\pi}
\end{equation}
Using values typical for our experiments, $\sigma_{\omega_{\mathrm{T}}} \approx 2 \pi \times 1 \text{kHz}$, compared to a typical value of $\omega_{\mathrm{T}} = 2 \pi \times 36 \text{kHz}$. Uncertainty from jitter is thus fairly small, but generally on the same order of magnitude as fitting error in $\omega_{\text{fit}}$. Error propagation gives:
\begin{equation}
    \sigma_{ C_\textrm{exp} } = \sqrt{\left( \sigma_{\omega_{\text{fit}}}/\omega_{\mathrm{T}} \right)^2 + \left(  \left| C_\textrm{exp} \right| \sigma_{\omega_{\mathrm{T}}}/ \omega_{\mathrm{T}} \right)^2}
\end{equation}
Where $\sigma_{\omega_{\text{fit}}}$ is the uncertainty in the fit parameters extracted from the 66\% confidence intervals of the fit.

\subsection{Data Analysis for Berry Curvature}
In this section, we expand on the data analysis performed on the measured overlaps to extract the local Berry curvature. Consider a pair of spins starting at drive phases $\Bt$ and $\Bt + \dBt$. If time evolves by a small amount $\delta t$, then the geometric phase between the spins will increase by the small amount $\delta \Phi = \left| \dBt \times \vec{\Omega} \delta t  \right| (\mathscr{B}_1(\Bt)-\mathscr{B}_2(\Bt))$, where we assume $\delta t$ is small enough so that $\mathscr{B}_j$ is constant over the area enclosed by the small path. Since the ground and excited state Berry curvatures are equal and opposite, we have that:
\begin{equation}
    \label{eq.app.PhiToB}
    \frac{\delta \Phi}{\delta t} = 2\left| \dBt \times \vec{\Omega} \right| \mathscr{B}_1(\Bt)
\end{equation}
So, the derivative of the phase is proportional to the local Berry curvature. Since the overlap is related to the phase by $F(t) = \cos ^2 \left( \frac{\Phi(t)}{2} \right) $, we can relate the derivative of the overlap to the local Berry curvature:
\begin{equation}
    \label{eq.app.derivatives}
    \frac{dF(t)}{dt} = -\frac{1}{2}\sin(\Phi) \frac{d \Phi (t)}{dt} = -\sin(\Phi) \left| \dBt \times \vec{\Omega}   \right| \mathscr{B}_1(\Bt)
\end{equation}
We have assumed $P_1=P_2=\tfrac12$, as is the case in our experiments. Deviation from $P_1=P_2=\frac12$ due to imperfect initialization just results in an overall multiplicative constant.

Due to the chain rule, a factor $\sin(\Phi)$ appears which will modulate the Berry curvature depending on what the value of $\Phi$ happens to be at each drive phase. Squaring the derivative gives:
\begin{equation}
    \label{eq.app.derivativeSquared}
    \left( \frac{dF(t)}{dt} \right)^2 = \sin^2(\Phi) \left| \dBt \times \vec{\Omega}   \right|^2 \mathscr{B}_1^2(\Bt)
\end{equation}
If we average this quantity over $\Phi$, we could replace $\sin^2(\Phi)$ by its average value of $\frac{1}{2}$, giving a result directly proportional to the magnitude of the local curvature. Note that the sign of the curvature can not be determined since the form of the overlap, $F(t) = \cos ^2 \left( \frac{\Phi(t)}{2} \right) $, does not contain information about the sign of the phase $\Phi(t)$.

\begin{figure*}[ht]
	\centering
	\includegraphics[width= .9\textwidth]{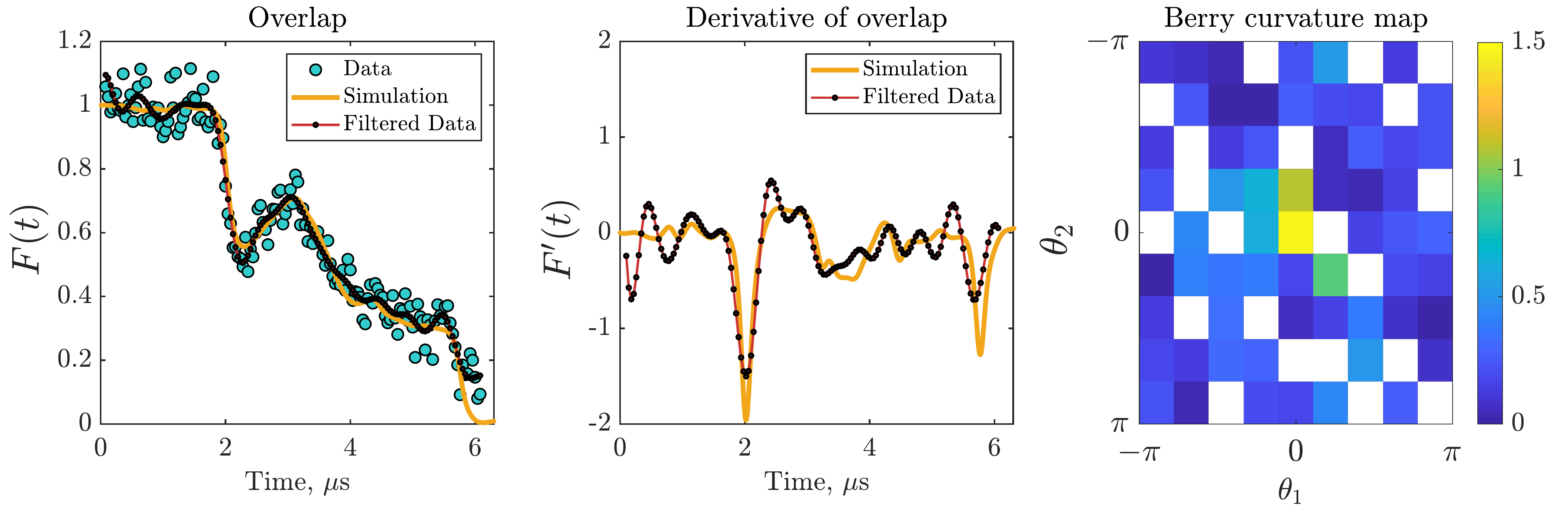}
	\caption{\textbf{Left:} Data and simulation for one representative data set used in mapping the Berry curvature. Blue dots are data, solid orange line is an \textit{a priori} simulation, and small red points joined by a red line are the data after smoothing with a low pass filter. Low pass filter is set with a bandpass of $2\pi \times 1$MHz bandpass and beginning and end are padded with repitions of the first and final data points. Filter has a stopband attentuation of 60dB and is implemented via the MatLAB function `lowpass'. \textbf{Middle:} Orange solid line is the derivative of the simulated overlap while red line with small markers is the derivative of the filtered data. Differentiation implemented using simple nearest neighbors difference. \textbf{Right:} Berry curvature constructed from the derivative of the smoothed data. White pixels are points that in this data set did not happen to be traversed by the trajectories. Since different data sets start at different initial phases, any given pixel is still covered by many different datasets. Hamiltonian parameters $\gamma B_0 = 2\pi \times 0.25 \uu{MHz}, \, m = 1.5, \, \vec{\Omega}=2\pi \left(0.5, 0.5\varphi \right) \uu{MHz}, \, \Bt_0 = \left(0, 3\pi/4 \right), \, \dBt_{0} = \left(\varphi/8, -1/8 \right)$}
	\label{fig.app.BerryCurve}
\end{figure*}

Suppose we want to use this method to measure the Berry curvature at a particular final drive phase, $\Bt_f$. If we start the drive phase at a range of initial values $\Bt_0$, then it will reach the final phase $\Bt_f$ in a variable amount of time. Since the geometric phase grows roughly linearly with time, the final geometric phase, $\Phi_f$ will also vary depending on the initial drive phase $\Bt_0$. Thus, by starting at different initial phases $\Bt_0$, we can produce a set of overlaps that end at the same final drive phase but will have accrued varying amounts of geometric phase. Averaging the squared derivative of the overlap evaluated at $\Bt_f$ over this data set will average out the modulating $\sin^2(\Phi)$ factor:
\begin{equation}
    \left< \left( \frac{dF(\Bt_f)}{dt} \right)^2 \right>_{\Bt_0} = \frac{1}{2} \left| \dBt \times \vec{\Omega}   \right|^2 \mathscr{B}_1^2(\Bt_f)
\end{equation}
Where the derivative of $F$ is evaluated at the time $t$ such that $\Bt = \Bt_f$, and the average $\left< \ldots \right>_{\Bt_0}$ indicates averaging over data sets with different initial phase $\Bt_0$.

The fidelity has technical noise that varies from point-to-point and is amplified by differentiation. The mean-squared averaging procedure results in an additive offset due to noise.
To mitigate the noise, we apply a low pass filter to the overlap data and then perform differentiation using $F'_i = (F_i - F_{i-1})/(t_i - t_{i-1})$, which we assign to the resampled time $t^{r}_i = (t_i + t_{i-1})/2$. The bandpass frequency of the filter is set to $1\text{MHz}$ since we expect the local variation will occur at the scale of the drive frequencies, $\sim \Omega$, and the larger of the drive frequencies is $2\pi \times 0.5\varphi \approx 0.8 \text{MHz} $, where $\varphi$ is the golden ratio. Near the critical point, we choose a higher bandpass frequency of $2.5 \text{MHz}$ to prevent broadening the observed overlap jumps. This results in a larger noise floor, as can be observed in Figure \ref{fig.berryMap} of the main text.

To determine the drive phase associated with each time we use $\Bt_{t} = \vec{\Omega}t + \Bt_0 $ since the drive frequency and initial phase are known. To account for the fact that the overlap is between two trajectories with different initial phase, we assign the derivative to the middle of the strip between the trajectories by taking $\Bt_{\mathrm{plot}} = \vec{\Omega} t_f + \Bt_0 + \frac{1}{2}\dBt $.

To perform the average over initial phase, we partition the Berry curvature into a grid of $N_b \times N_b$ pixels, and average together the square of the derivative of the overlap for all drive phases that fall into each pixel. A minimum pixel size is set by the width of the trajectories, $ \left| \dBt \right|/2 \pi \sim 0.04 $. This implies a maximum number of pixels of about $N_b=25$, but we use $N_b=9$ for convenience since many data points for each pixel must be collected to average out the the $\sin^2(\Phi)$ factor.

Figure \ref{fig.app.BerryCurve} shows a representative data set. The right panel shows the raw overlap data, filtered data, and simulated overlap, the middle panel shows the measured and simulated derivative, and the right panel shows the berry curvature for this individual data set. Note that the chosen bandpass filter does a good job preserving the shape of the curve in regions where $\Phi$ changes rapidly, but still has spurious fluctuations in regions where it changes more slowly.

\section{Supplemental Measurements}

\subsection{Overlap oscillations with commensurate driving frequencies}

Generating incommensurate frequencies can be difficult as signal generating hardware often produces tones at harmonics of a fundamental frequency, meaning all drives will be commensurate. If we approximate the irrational ratio of the drive frequencies by the sequence of rational numbers $p_n/q_n \approx \Omega_2/\Omega_1$, how large does $n$ need to be to observe the topological properties investigated in this work? In this section we present data and simulations that show that $n$ can be as small as $6$.

If the drives are commensurate, then the Hamiltonian is periodic with period $T = p_n \Omega_1 = q_n \Omega_2$ and a pair of drive phases will not cover the entire synthetic Brillioun zone before returning to their starting values. The Berry curvature is not guaranteed to integrate to an integer along such a trajectory and the overlap oscillations may occur at arbitrary frequencies. Since each trajectory only samples part of the Berry curvature, trajectories starting from different initial drive phases may result in different oscillation frequencies. Conversely, as $q_n$ becomes larger, the ratio of drive frequencies will more closely approximate an irrational number and the trajectories will sample the synthetic Brillioun zone more densely. This leads to less variation of the overlap oscillation frequency with initial phase as all values approach the expected quantized value.

\begin{figure}[ht]
	\centering
	\includegraphics[width= .5\textwidth]{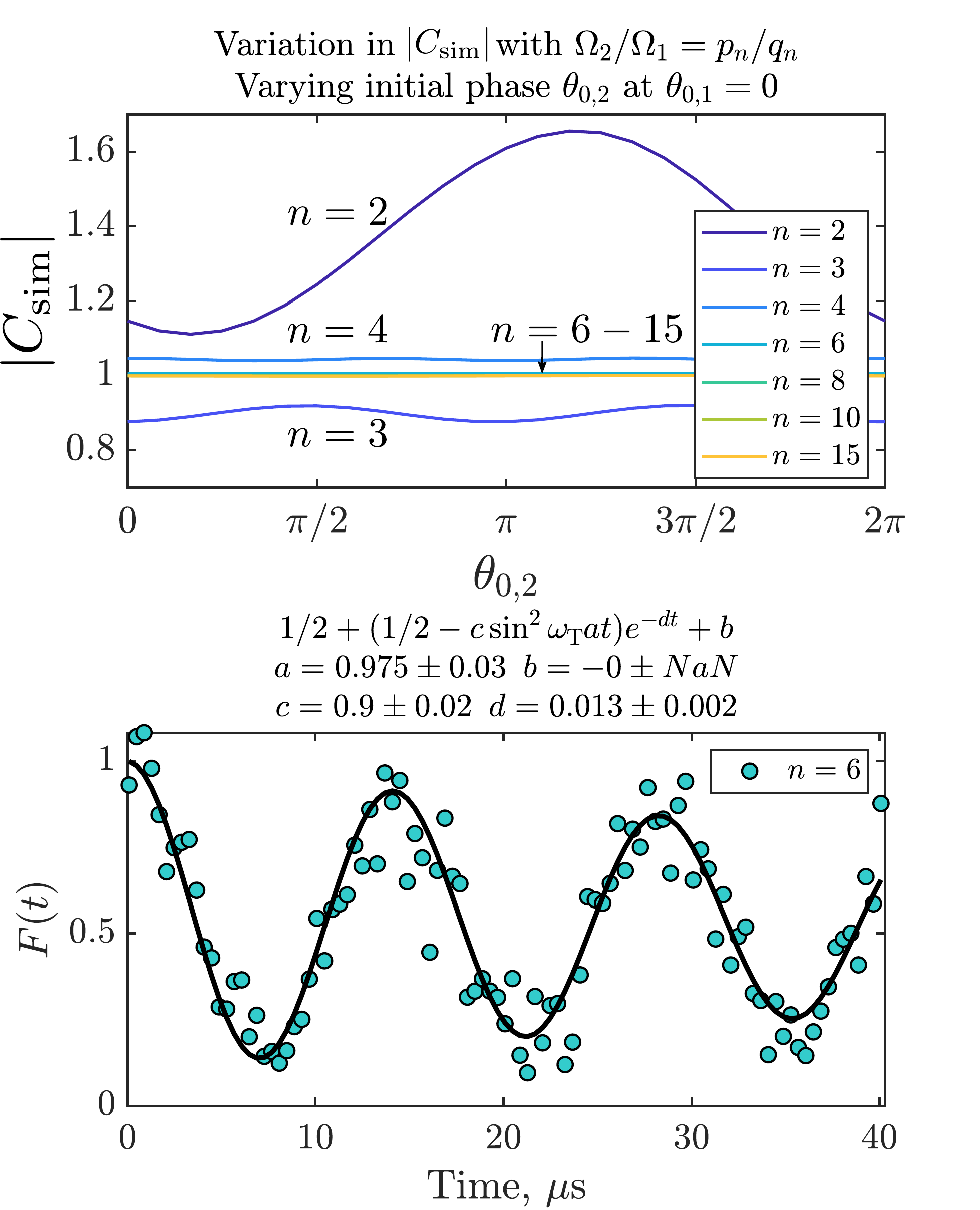}
	\caption{\textbf{Top:} Effective Chern number extracted from simulated overlap oscillations as a function of initial drive phase at various approximations to $\Omega_2/\Omega_1 \approx \varphi \approx F_{n+1}/F_n$ for Fibonacci numbers $F_{n}$. Hamiltonian parameters: $T_{tot}= 1000\mu \text{s}, \,  \gamma B_0 = 2\pi \times 0.25 \uu{MHz}, \, m = 1,  \, \vec{\Omega}=2\pi \left(0.5, 0.5 F_{n+1}/F_n \right) \uu{MHz}, \, \Bt_0 = \left(0, \pi/2 \right), \, \dBt_0 = \left((F_{n+1}/F_n)/8, -1/8 \right)$. \textbf{Bottom:} measurements (blue data points) and fits (black line) to the overlap with commensurate frequencies where $\Omega_2/\Omega_1 = F_{7}/F_6 = 13/8$. Fit parameter $b$ fixed at 0 to prevent the regression optimizer from finding spurious minima with unrealistic fit values. $\omega_{\mathrm{T}} =  \frac{\left| \vec{\Omega} \times \dBt_0  \right|}{2 \pi}$ scales the frequency so that the fit value of $a$ is the magnitude of the Chern number. Parameters are the same as the simulation parameters for the case $n=6$}
	\label{fig.app.commensurate}
\end{figure}

In the top panel of Figure \ref{fig.app.commensurate}, we perform simulations of the overlap oscillations and report the effective Chern number of the oscillations as a function of the initial drive phase. We show several curves with increasing $n$ where the ratio of the drives more closely approximates the irrational number $\varphi$, the golden ratio, so that approximations are given by $\Omega_2/\Omega_1 = F_{n+1}/F_n$, where $F_n$ is the $n^{th}$ Fibonacci number. Low values of $n$ are bad approximations of $\varphi$ and the extracted Chern number varies with initial phase and is not an integer. As $n$ increases, the behavior improves and even by only $n=6$ where $\varphi \approx 13/8$ The Chern number is nearly 1 and there is very little variation with the initial drive phase. This indicates this is a sufficient approximation to see the topological behavior at the level of precision available in our experiments.

To verify this, we perform an experiment in the commensurate regime with $n=6$. As the data in the bottom panel of Figure \ref{fig.app.commensurate} shows, the overlap oscillations are clear and we measure $\left| C_\textrm{exp} \right| = 0.98 \pm 0.03$, consistent with a Chern number of 1. To the level of precision available in our data, it does not appear necessary that the drives be incommensurate to a high degree of precision.

\subsection{Effect of spin-echo on measurements}

In this section we investigate the effect of the spin echo on the measurement results. To evaluate how well the spin-echo performs at decreasing decoherence, we measure the overlap in the topological regime for two experiments with identical parameters except for the presence or absence of the echo pulse. Note that we only transform the fields as $B_y \rightarrow -B_y, \, B_z \rightarrow -B_z$ for the second half of the evolution when applying the echo pulse. 

\begin{figure}[ht]
	\centering
	\includegraphics[width= .5\textwidth]{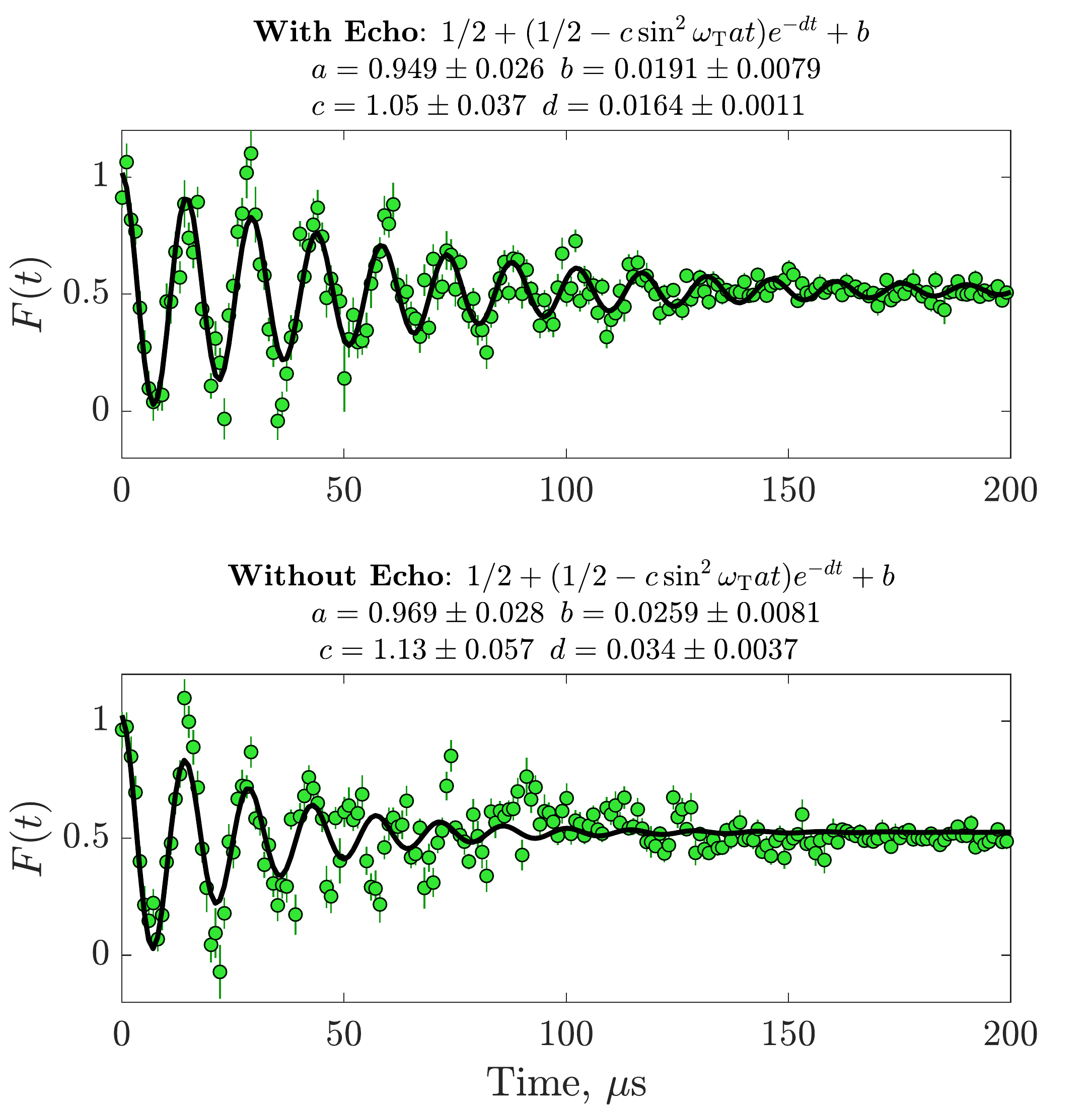}
	\caption{Measurements of and fits to the overlap in the topological phase with and without the echo pulse. The regression model is displayed in the title and error bars on data points are one standard error. \textbf{Top:} green circles are data points and black line is fit results when using an echo pulse. \textbf{Bottom:} green circles are data points and black line is fit results when \textbf{not} using an echo pulse. $\omega_{\mathrm{T}} =  \frac{\left| \vec{\Omega} \times \dBt_0  \right|}{2 \pi}$ scales the frequency so that the fit value of $a$ is the magnitude of the Chern number. Hamiltonian parameters: $\gamma B_0 = 2\pi \times 0.25 \uu{MHz}, \, r = 1,  \, \vec{\Omega}=2\pi \left(0.5, 0.5\varphi \right) \uu{MHz}, \, \Bt_0 = \left(0, \pi/2 \right), \, \dBt_0 = \left(\varphi/8, -1/8 \right)$}
	\label{fig.app.echos}
\end{figure}

The results in Figure \ref{fig.app.echos} show that the observed decay times are $\tau =  61  \pm 4\, \mu \text{s}$ for the run with the spin echo and $\tau = 29 \pm 3 \, \mu \text{s}$ for the run without. The measurement with the spin echo exhibits much clearer oscillations, especially around $60-100 \mu \text{s}$. The spin echo is able to extend the lifetime of the topological behavior to a time consistent with the $T_2$ of $125\pm 7 \, \mu \text{s}$. Since the overlap scales as $F \sim \langle i \rangle^2$, and spin-echo measures the coherence of $\langle i \rangle$, we expect the coherence time to be half as long, as observed. This suggests the topological behavior is amenable to dynamical decoupling methods and compatible with the unavoidable pulse errors that accompany them. Additionally, the $T_2^*$ measured by a detuned Ramsey experiment is $ T_2^* = 6.4 \pm 0.5 \, \mu \text{s}$, significantly shorter than the observed decay time for the half-BHZ model even without the spin echo. This indicates that the half-BHZ Hamiltonian by itself may perform some kind of decoupling effect.

\begin{figure}[ht]
	\centering
	\includegraphics[width= .5\textwidth]{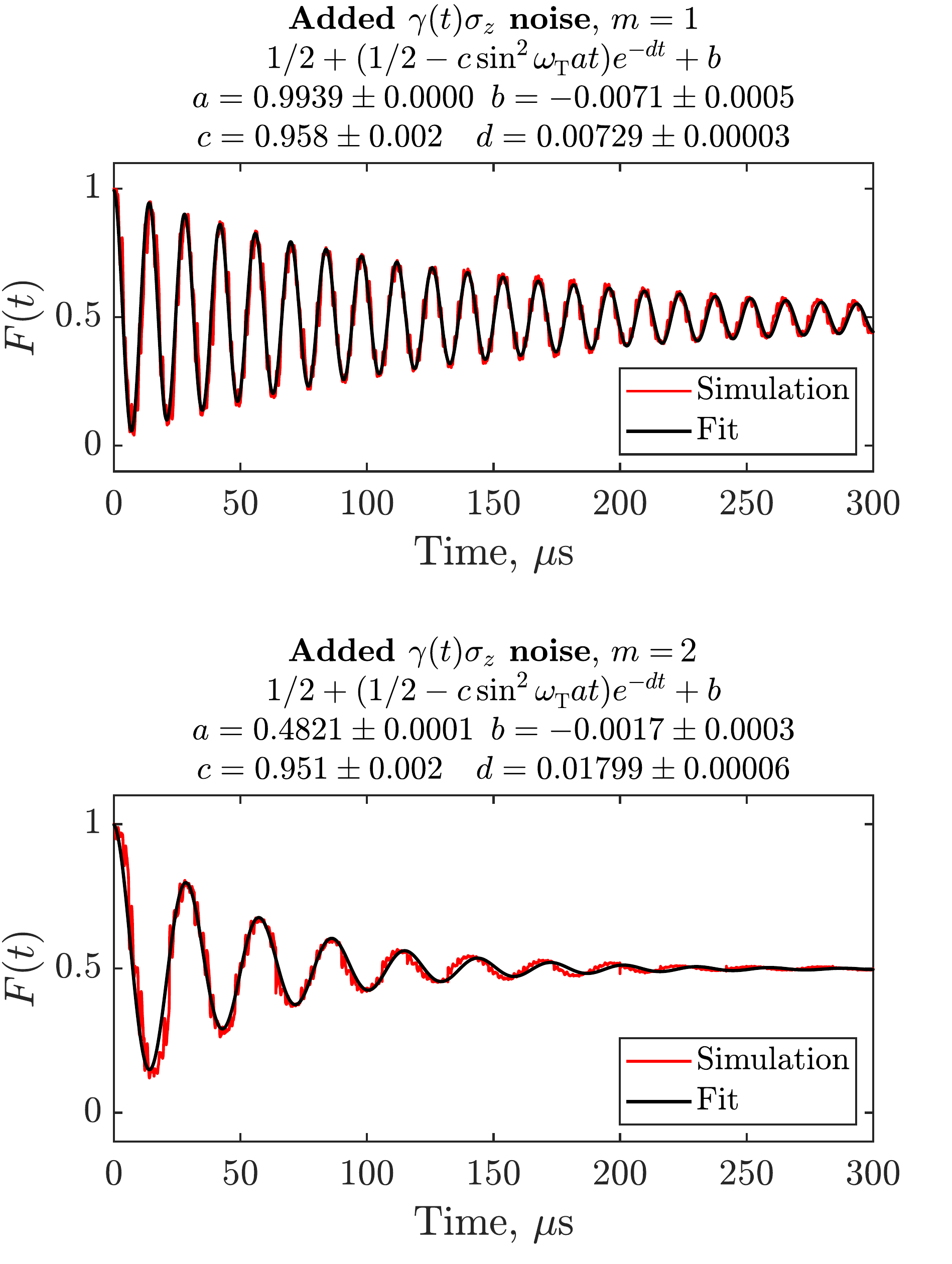}
	\caption{Simulation of the overlap oscillations with additive band limited white noise. Simulations are performed by numerically integrating the Hamiltonian $H_{\mathrm{CD}}+\gamma(t)\sigma_z$, where $\gamma(t)$ is band-limited white noise, to get the state as a function of time followed by applying pulses and measuring the population of $\ket{+z}$ to simulate the measurement procedure described above. This is repeated 500 times with different realizations of the noise and the measurements $M_{\pm j}(\tau)$ are averaged together. The overlap is then constructed as detailed above. $\omega_{\mathrm{T}} =  \frac{\left| \vec{\Omega} \times \dBt_0  \right|}{2 \pi}$ scales the frequency so that the fit value of $a$ is the magnitude of the Chern number. Parameters: $\gamma B_0 = 2\pi \times 0.25 \uu{MHz}, \, \vec{\Omega}=2\pi \left(0.5, 0.5\varphi \right) \uu{MHz}, \, \Bt_0 = \left(0, \pi/2 \right), \, \dBt_0 = \left(\varphi/8, -1/8 \right)$, $\gamma(t)$ has root-mean-square amplitude $2\pi \times 0.006 \uu{MHz}$ and bandwidth $2\pi \times 0.1 \uu{MHz}$. \textbf{Top:} Topological regime, $m=1$; \textbf{Bottom:} Critical point, $m=2$}
	\label{fig.app.transitionDecay}
\end{figure}

Near the critical point at $m = 2$ the observed decay rate is larger, as can be seen in Figure \ref{fig.divergence.fid}(a) of the main text, with values around $50 \, \mu \text{s}$. This is expected since, as discussed in appendix~\ref{app.VCD}, near the critical point the counter-diabatic potential is likely to be imperfect. Numerical simulation of the Schr{\"o}dinger equation with added band-limited white noise corroborates the observation that decay is faster near the critical point, as shown in Figure \ref{fig.app.transitionDecay}. 

\subsection{Overlap oscillations without counterdiabatic driving}
Throughout this work, we add a counterdiabatic potential to the half-BHZ Hamiltonian which prevents excitations between the quasienergy eigenstates and allows us to work outside the adiabatic regime. However, the counterdiabatic potential is not a requirement for the topological behavior and the same method can be used to measure the Chern number without it. Here, we show the results for experiments without the counterdiabatic potential.

\begin{figure}[ht]
	\centering
	\includegraphics[width= 0.46\textwidth]{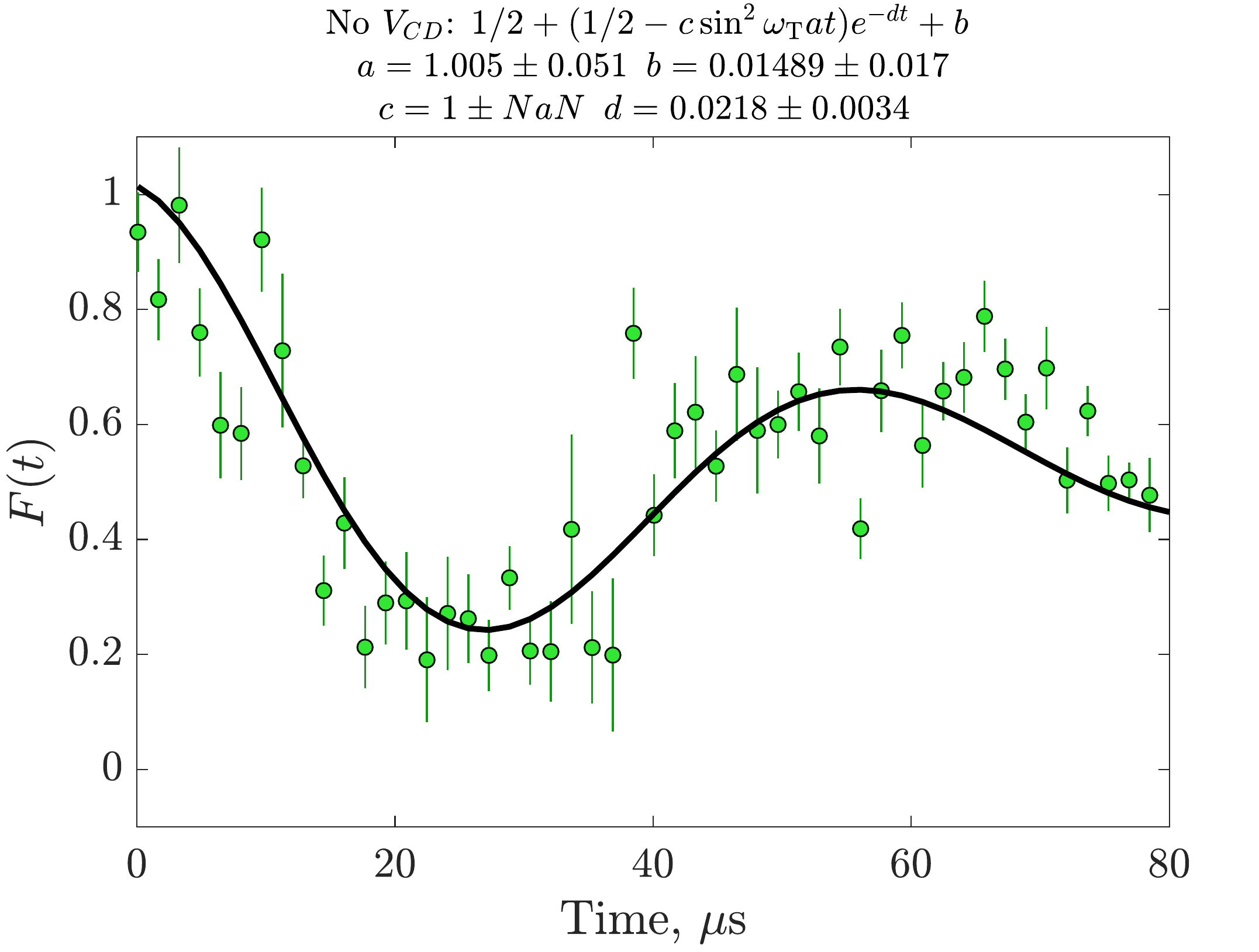}
	\caption{Measurement of (green circles) and fit to (black line) the overlap oscillations in the topological regime without an added counterdiabatic potential. Error bars on data points represent 1 standard deviation and uncertainties on fit parameters are 1 standard error. Uncertainty of NaN indicates the parameter was fixed to prevent overfitting to spurious minima with highly unphysical values. $\omega_{\mathrm{T}} =  \frac{\left| \vec{\Omega} \times \dBt_0  \right|}{2 \pi}$ scales the frequency so that the fit value of $a$ is the magnitude of the Chern number. Hamiltonian parameters: $\gamma B_0 = 2\pi \times 1 \uu{MHz}, \, m = 1,  \, \vec{\Omega}=2\pi \left(0.3, 0.3\varphi \right) \uu{MHz}, \, \Bt_0 = \left(\pi, \pi/2 \right), \, \dBt_0 = \left(\varphi/20, -1/20 \right)$}
	\label{fig.app.noCD}
\end{figure}

To work near the adiabatic regime, we need $\gamma B_0 \gg \Omega_1, \Omega_2$ so that the rate of change of the Hamiltonian is always small compared to the instantaneous gap. Increasing $B_0$ also increases the size of fluctuations in the overlap due to the dynamical phase, since it is only approximately cancelled by our measurement procedure. Therefore, we choose parameters with larger $B_0$, smaller $\vec{\Omega}$, smaller $\dBt_0$, and longer measurement duration $\tau$, resulting in fewer oscillations for a given evolution time.

Figure \ref{fig.app.noCD} shows the results in the topological regime without a counterdiabatic potential. The parameter values resulted in a much lower topological frequency and the decay rate was higher due to the additional heating from the drives. Despite these issues, we still observe the expected oscillation frequency corresponding to a Chern number of $\left| C_\textrm{exp} \right| = 1.01 \pm 0.05$. Thus, counterdiabatic driving is helpful for observing the topological oscillations, but not necessary.

\subsection{Experimental estimation of overlap decline from decoherence}
The Chern number is zero in the trivial regime, so there should be no decline in the overlap due to geometric phase. We observe a small decline in the overlap which we conservatively assume was caused by a non-zero Chern number in order to place an upper bound on it. However, it is more likely this decline is caused be decoherence, which will decrease the magnitude of the expectation values and hence the overlap.

\begin{figure}[ht]
	\centering
	\includegraphics[width= .5\textwidth]{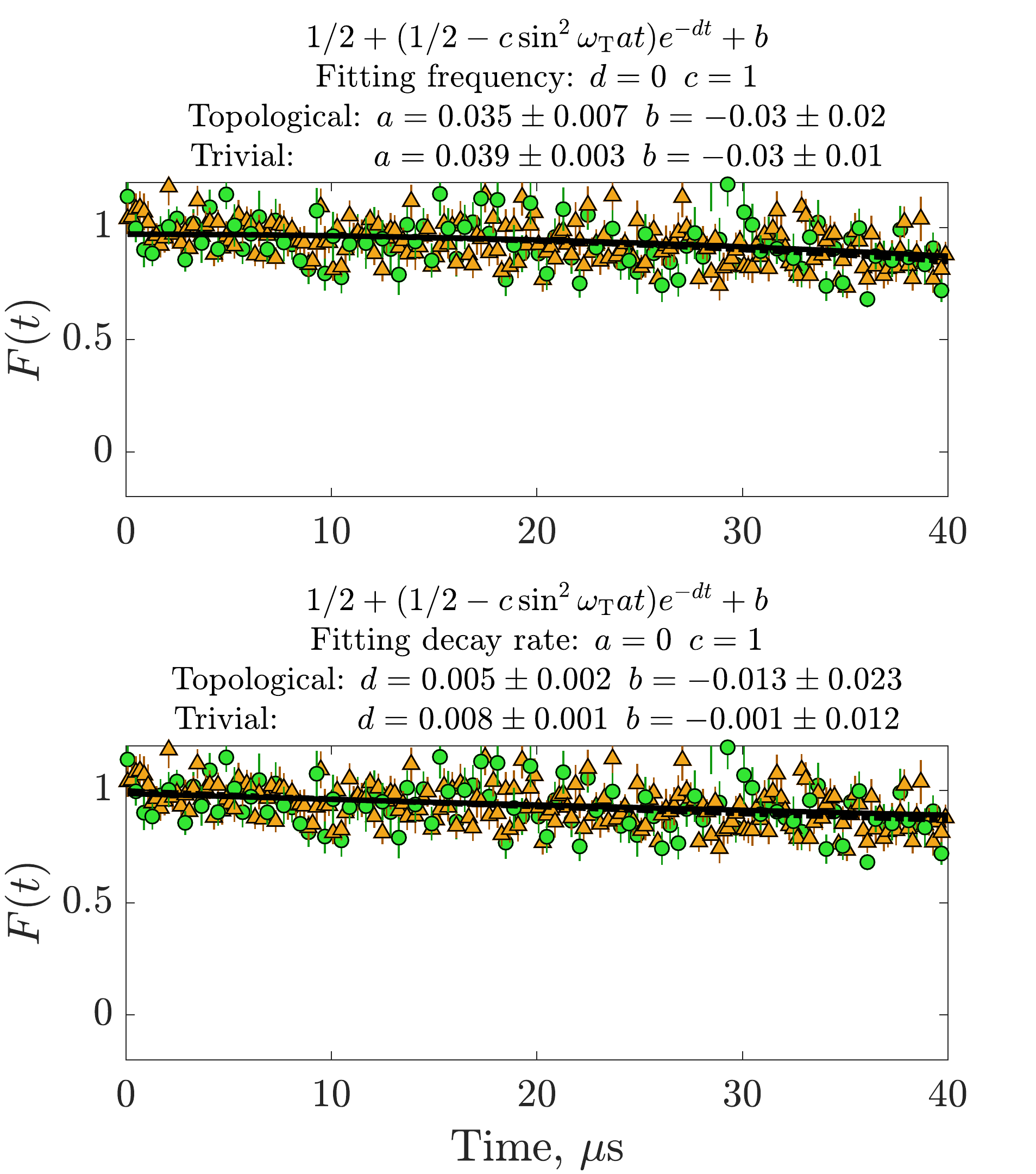}
	\caption{Measurements of and fits to the overlap in the trivial regime, $m = 3$, with $\dBt_0 \perp \vec{\Omega}$ (orange triangles, dot-dashed line) and in the topological regime, $m = 1$, with $\dBt_0 \parallel \vec{\Omega}$ (green circle, solid lines). Top and bottom panels show the same data but fit different parameters in the overlap model to investigate scenarios where the decline in the overlap is caused entirely by a non-zero Chern number (top) or by decoherence (bottom). Error bars on data points are 1 standard devitation and error bars on fit parameters are 1 standard error. Hamiltonian parameters: $\gamma B_0 = 2\pi \times 0.25 \uu{MHz}, \, m = 1, \, 3  \, \vec{\Omega}=2\pi \left(0.5, 0.5\varphi \right) \uu{MHz}, \, \Bt_0 = \left(0, \pi/2 \right), \, \dBt_{0, \, \parallel} = \left(1/8, \varphi/8 \right), \, \dBt_{0, \, \bot} = \left(\varphi/8, -1/8 \right)$. $\omega_{\mathrm{T}} =  \frac{\left| \vec{\Omega} \times \dBt_{0, \bot}  \right|}{2 \pi}$ scales the frequency so that the fit value of $a$ is the magnitude of the Chern number.}
	\label{fig.app.parallel}
\end{figure}

To test this, we perform measurements of the overlap in the topological regime $(m=1)$, but where we choose $\dBt$ to be parallel to $\vec{\Omega}$ so that the area between the two trajectories is zero and the geometric phase is therefore zero regardless of the Chern number. The results in Figure \ref{fig.app.parallel} (green circles) show there is a small decline in the overlap which must be caused by decoherence. These results are consistent with the results in the trivial regime with $\dBt$ perpendicular to $\vec{\Omega}$ (orange triangle). Thus, the decline in the overlap in the trivial regime can likely be attributed to decoherence and not to a non-zero Chern number. For completeness, we show fits to models where the decline is assumed to be due to geometric phase and decoherence; in both cases the results are consistent.

\subsection{Measured Chern number as function of initial phase}
\begin{figure}[ht]
	\centering
	\includegraphics[width= .5\textwidth]{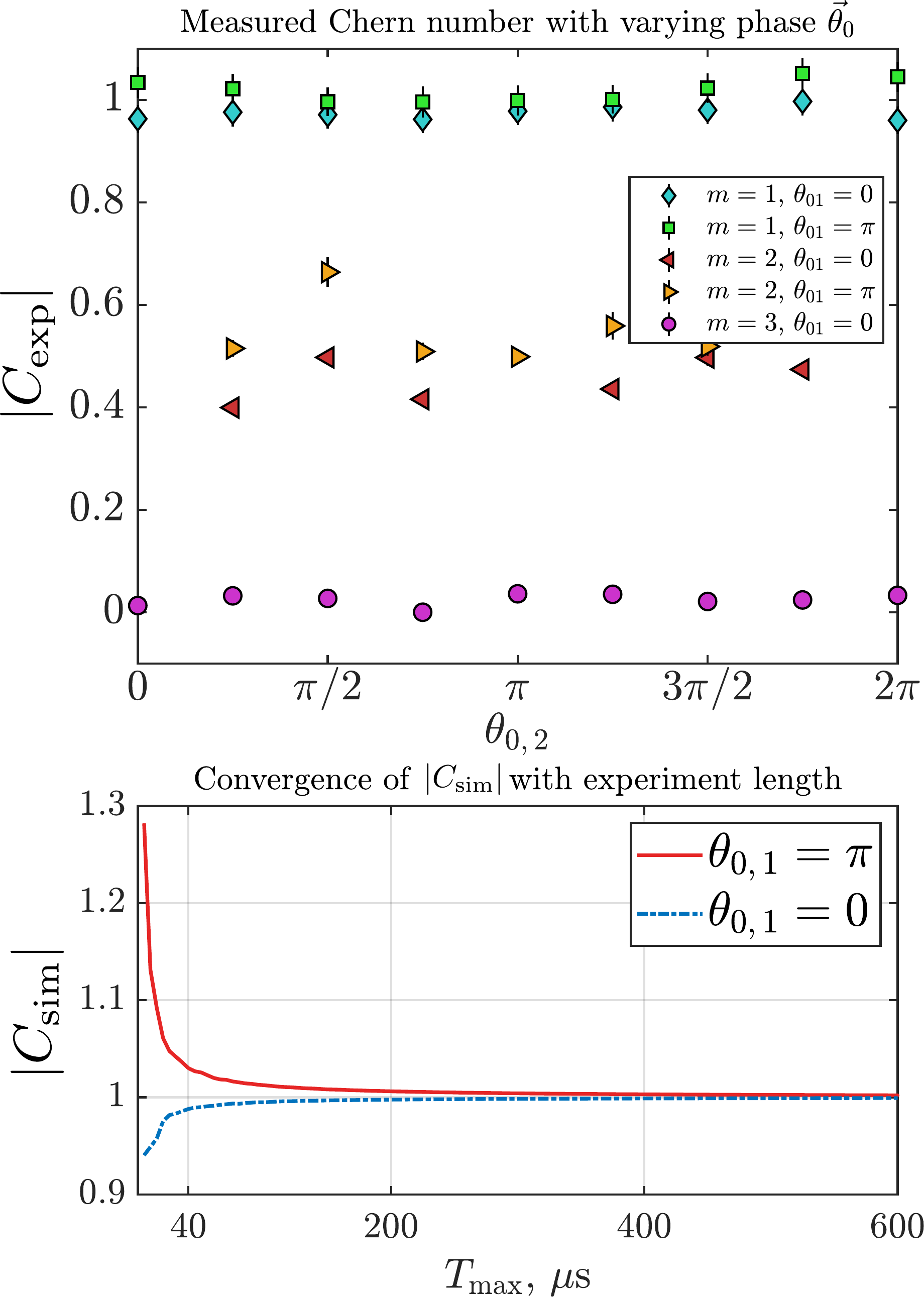}
	\caption{\textbf{Top:} Measurements of the Chern number as a function of initial phase $\Bt_0 = \left( \theta_{01}, \theta_{02} \right)$ in the topological (green squares and cyan diamonds), critical (orand and red triangles), and trivial (purple circle) regimes. The Chern number varies randomly with $\theta_{02}$ due to scatter in the data, but has a systematic difference for $\theta_{01}$ as explained. Hamiltonian parameters $\gamma B_0 = 2\pi\times 0.25 \uu{MHz}, \, m = 1, \, 2, \, 3, \,  \, \vec{\Omega}=2\pi \left(0.5, 0.5\varphi \right) \uu{MHz}, \, \dBt_{0} = \left(\varphi/8, -1/8 \right) T_{\mathrm{\max}} = 40 \mu$s. \textbf{Bottom:} Chern number extracted from simulated overlap as a function of the total length of the simulation from $5\mu$s to $1000\mu$s. Parameters are the same as the top panel with $\theta_{02} = 0$ and $\theta_{01} = 0$ (dot-dashed blue curve) and $\theta_{01} = \pi$ (solid red curve)}
	\label{fig.app.phaseVary}
\end{figure}

In the main text, most of the data we present is for a single value of the initial phase, $\Bt_0 = \left(0, \, \pi/2 \right)$, chosen because it is easy to prepare the spin in the state $\ket{\uparrow_z}$, which is an even superposition of the initial instantaneous eigenstates. Here we show that our results do not depend on the choice of initial phase, up to small systematic effects that can be accounted for. The top panel of Figure \ref{fig.app.phaseVary} shows the Chern number measured in each regime as a function of the initial phase $\theta_{0,2}$, for which $\ket{\uparrow_z}$ is still an even superposition of the initial eigenstates. We see that in the topological and trivial regimes there is very little change in the Chern number as the initial phase varies.

However, in the topological regime there is a small systematic effect observable: the results for $\theta_{0,1} = \pi $ are consistently slightly larger than for $\theta_{0,1} = 0$. This occurs because the local Berry curvature is approximately constant along the lines $\theta_{0,1} = 0, \, \theta_{0,2} = 0$ and larger than at phases away from these lines, as shown in Figure \ref{fig.berryMap} in the main text. Since the phase in the overlap oscillations is the integrated Berry curvature, starting along $\theta_{0,1} = 0$ where the curvature is high means that the first period of the oscillations will appear slightly longer than is typical because half of the region with higher curvature is missed. Conversely, starting well away from $\theta_{0,1} = 0$ means it is likely the entire region of higher curvature will occur in the first period, leading to a slightly shorter period.

This transient effect should become unimportant for longer experiments as the regions of higher and lower curvature occur uniformly throughout the oscillations, and the results should converge to the Chern number for all initial phases. As the simulations in the bottom panel of Figure \ref{fig.app.phaseVary} show, as the experiment duration becomes longer, the transient effect indeed dies away and the results converge to $\left| C \right| = 1$ for both choices of $\theta_{0,1}$. The experiments in the top panel are performed for $T_{\mathrm{max}}=40 \mu$s where the systematic effect is still predicted to be noticeable, accounting for the observed systematic difference in our data. This effect also applies to the data presented in the main text, particularly in Figure \ref{fig.divergence.fid} where our data systematically, if only slightly, underestimates the Chern number.

At the critical point of $m = 2$, the same effect occurs, as the data at $\theta_{0,1} = \pi $ generally have a slightly larger measured Chern number than at $\theta_{0,1} = 0$. In addition, there is more spread in the data. First, half as many periods occur in the same experiment duration so fitting the overlap oscillations will be less robust and the fit parameters are expected to have higher variance. Second, as discussed above, decoherence effects the critical point more, which dampens the oscillations and again will lead to more variance in the fit since the oscillations are not as well resolved. Finally, as mentioned above in appendix \ref{app.VCD}, at the critical point the counterdiabatic potential is likely to be imperfect, allowing excitations to occur. 

\end{document}